\title{\textbf{Searching Polyhedra\\
by Rotating Half-Planes}}
\author{
        \textsc{Giovanni Viglietta}\\
        Department of Computer Science\\
        University of Pisa\\
        \normalsize
        \texttt{viglietta@gmail.com}
}
\date{\today}

\documentclass[12pt,a4paper]{article}
\usepackage{amsfonts}
\usepackage{amsmath}
\usepackage{amsthm}
\usepackage{amssymb}
\usepackage{hyperref}

\usepackage{graphicx}
\usepackage{epsfig}
\usepackage{subfigure}
\usepackage{psfrag}
\usepackage[font=small,labelfont=bf,margin=10pt]{caption}

\newtheoremstyle{mystyle}{}{}{\itshape}{}{\bfseries}{.}{5 pt}{\thmname{#1}\thmnumber{ #2}\thmnote{ {\bfseries(#3)}}}
\theoremstyle{mystyle}
\newtheorem{theorem}{Theorem}

\newtheorem{corollary}[theorem]{Corollary}
\newtheorem{proposition}[theorem]{Proposition}
\newtheorem{lemma}{Lemma}

\newtheorem{definition}{Definition}
\numberwithin{equation}{section}

\newcommand{\R}{\mathbb{R}}
\newcommand{\complexityset}[1]{\textsc{#1}}
\renewcommand{\P}{\complexityset{P}}
\renewcommand{\L}{\complexityset{L}}
\newcommand{\NP}{\complexityset{NP}}
\newcommand{\PTAS}{\complexityset{PTAS}}
\newcommand{\PSPACE}{\complexityset{PSPACE}}
\newcommand{\NPSPACE}{\complexityset{NPSPACE}}
\newcommand{\SSP}{\textsc{SSP}}
\newcommand{\TSSP}{\textsc{3SSP}}
\newcommand{\TRSSP}{\textsc{3PSSP}}
\newcommand{\ART}{\textsc{Art Gallery Problem}}
\newcommand{\TSAT}{\textsc{3SAT}}
\newcommand{\EENCL}{\textsc{EE-NCL}}
\newcommand{\EEANCL}{\textsc{EE-ANCL}}

\begin{document}

\maketitle

\begin{abstract}
The \textsc{Searchlight Scheduling Problem} was first studied in 2D polygons, where the goal is for point guards in fixed positions to rotate searchlights to catch an evasive intruder. Here the problem is extended to 3D polyhedra, with the guards now boundary segments who rotate half-planes of illumination.

After carefully detailing the 3D model, several results are established. The first is a nearly direct extension of the planar one-way sweep strategy using what we call \emph{exhaustive} guards, a generalization that succeeds despite there being no well-defined notion in 3D of planar \textquotedblleft clockwise rotation\textquotedblright. Next follow two results: every polyhedron with $r>0$ reflex edges can be searched by at most $r^2$ suitably placed guards, whereas just $r$ guards suffice if the polyhedron is orthogonal. (Minimizing the number of guards to search a given polyhedron is easily seen to be \NP-hard.) Finally we show that deciding whether a given set of guards has a successful search schedule is strongly \NP-hard, and that deciding if a given target area is searchable at all is strongly \PSPACE-hard, even for orthogonal polyhedra. A number of peripheral results are proved en route to these central theorems, and several open problems remain for future work.
\end{abstract}

\section{Introduction}
\subsubsection*{Previous work}
The \textsc{Searchlight Scheduling Problem} (\SSP) was first studied in \cite{search1} as a search problem in simple polygons, where some stationary guards are tasked to locate an evasive, moving intruder by hitting him with searchlights. Each guard carries a searchlight, modeled as a 1-dimensional ray that can be continuously rotated, while the intruder runs unpredictably and with unbounded speed, trying to avoid the searchlights. Since the guards cannot know the position of the intruder until they catch him in their lights, the movements of the searchlights must follow a fixed \emph{schedule}, which should guarantee that the intruder is caught in finite time, regardless of the path he decides to take. The search takes place in a polygonal region, whose sides act as obstacles both for the intruder's movements and for the guards' searchlights. In a way, the polygonal boundary benefits the intruder, who can hide behind corners and avoid scanning searchlights. But it can also turn into a cul-de-sac, if the guards manage to force the intruder into an enclosed area from which he cannot escape.

\begin{figure}[ht]
\centering
\subfigure[]{\includegraphics[width=0.35\linewidth]{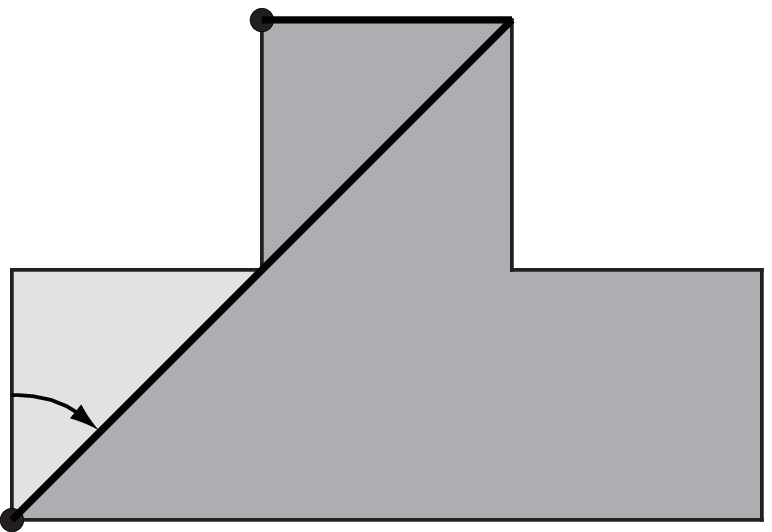}}\qquad \qquad
\subfigure[]{\includegraphics[width=0.35\linewidth]{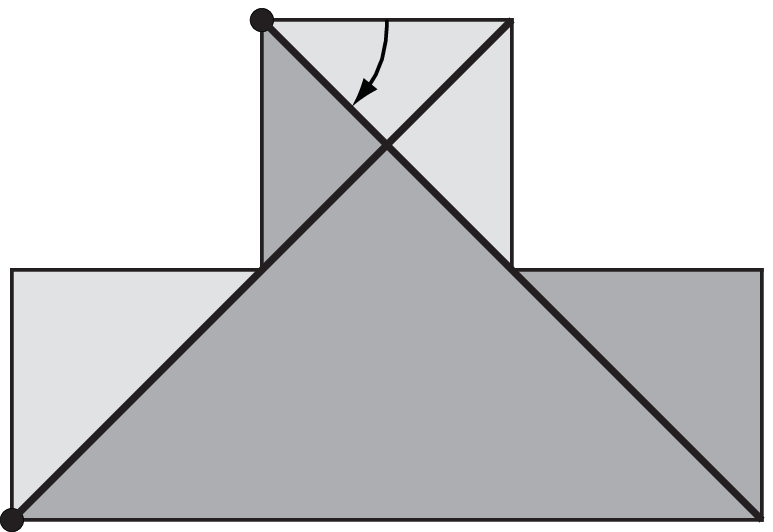}}\\ \vspace{0.5cm}
\subfigure[]{\includegraphics[width=0.35\linewidth]{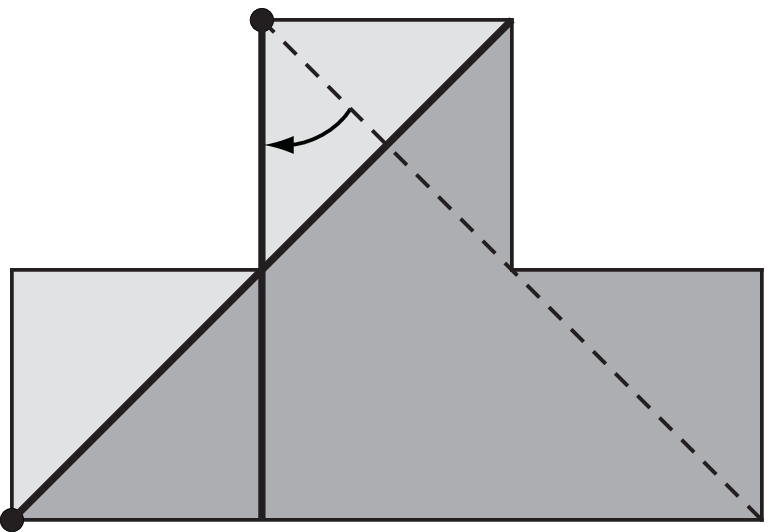}}\qquad \qquad
\subfigure[]{\includegraphics[width=0.35\linewidth]{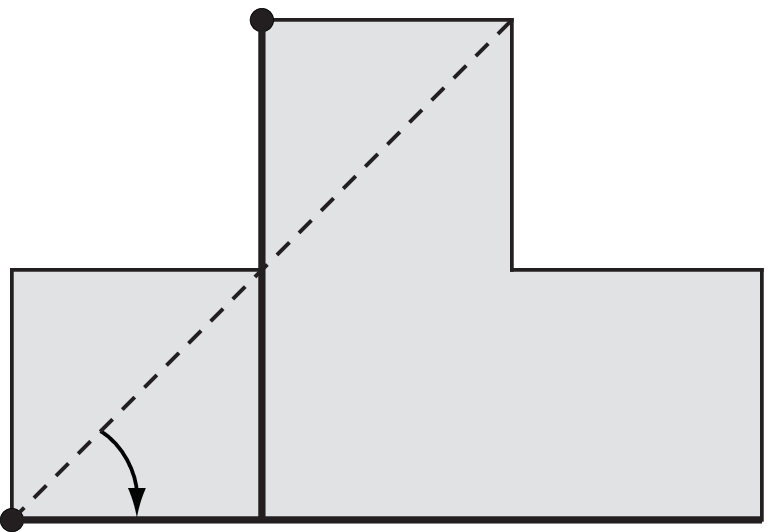}}
\caption{A search schedule for two guards in a polygon. At each stage, the dark area is still \textquotedblleft contaminated\textquotedblright, the lighter areas have been cleared.}
\label{fig:1}
\end{figure}

Thus \SSP\ is the problem of deciding if  there exists a successful search schedule for a given finite set of guards in a given simple polygon. Figure~\ref{fig:1} shows an instance of \SSP\ with a successful schedule.

\begin{sloppypar}We will now briefly review some well-known results pertaining \SSP\ in 2-dimensional polygons, before summarizing our results for the 3-dimensional model.\end{sloppypar}

\begin{sloppypar}A trivial necessary condition for searchability is that the guard positions should guarantee that no point in the polygon is invisible to all guards. In other words, the guards should at least solve the \ART\ in the given polygon (see \cite{art}), otherwise the intruder could sit at an uncovered point and never be discovered.\end{sloppypar}

Another simple necessary condition is that every guard lying in the interior of the polygon (thus not on the boundary) should be visible to at least one other guard. Without this, the intruder could remain in a neighborhood of a guard and just avoid its rotating searchlight.

Several sufficient conditions for searchability are detailed in \cite{search1}, all employing a general search algorithm called the \emph{one-way sweep strategy}. Most notably, if all the guards lie on a simple polygon's boundary and collectively see its whole interior, then they also have a successful search schedule.

Concerning the problem of minimizing the number of guards to search a given polygon, \cite{search1} contains a characterization of the simple polygons that are searchable by one or two suitably placed guards. A similar characterization for three guards was also found by the same authors, but never published (\cite{mark}). On the other hand, \cite{search2} contains some upper bounds on the minimum number of guards required to search a polygon (possibly with holes) as a function of the number of guards needed to solve the \ART\ in the same polygon.

\begin{figure}[ht]
\centering
\psfrag{a}{$\ell_1$}
\psfrag{b}{$\ell_2$}
\subfigure[]{\includegraphics[width=0.29\linewidth]{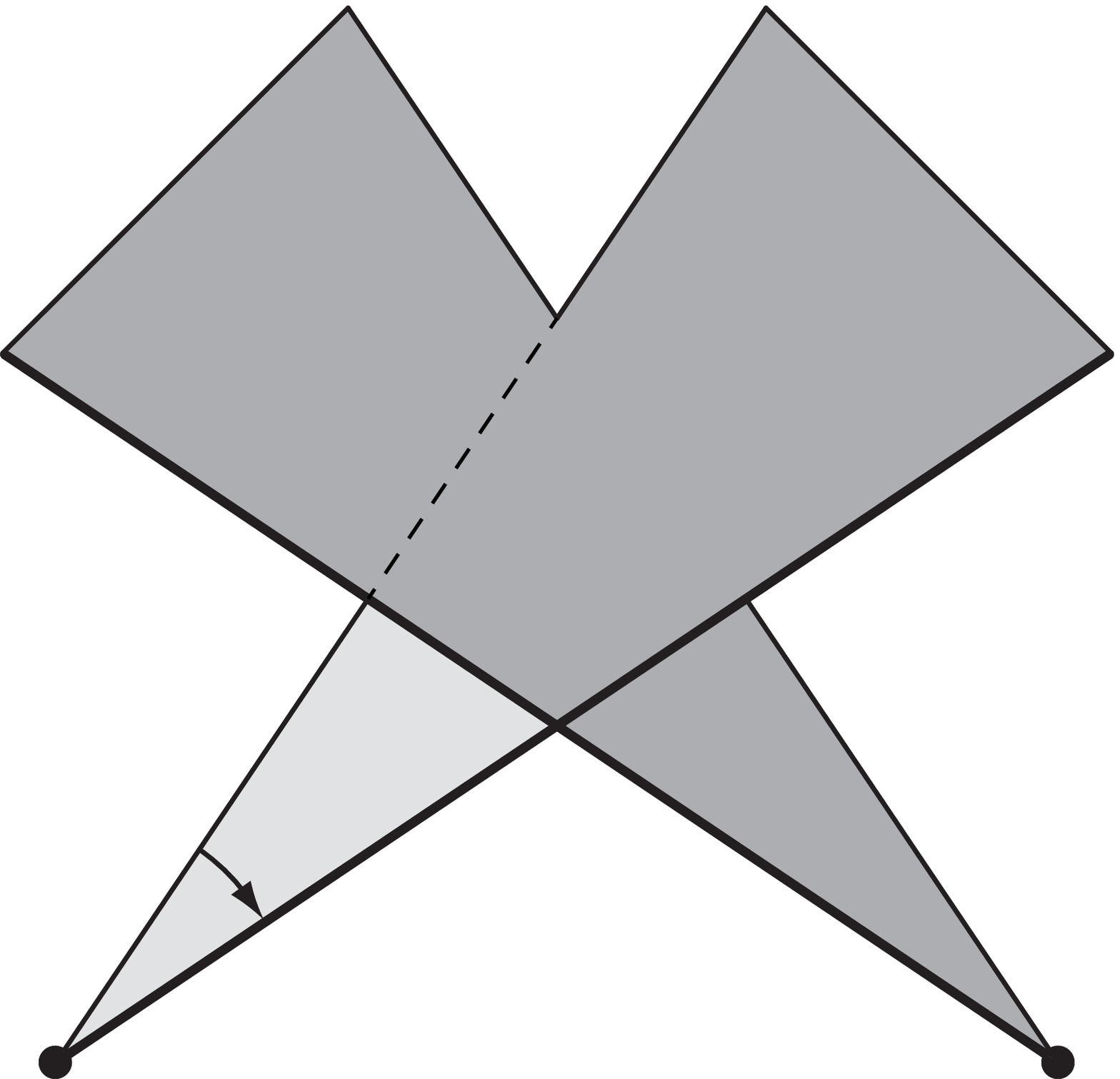}}\quad \ 
\subfigure[]{\includegraphics[width=0.29\linewidth]{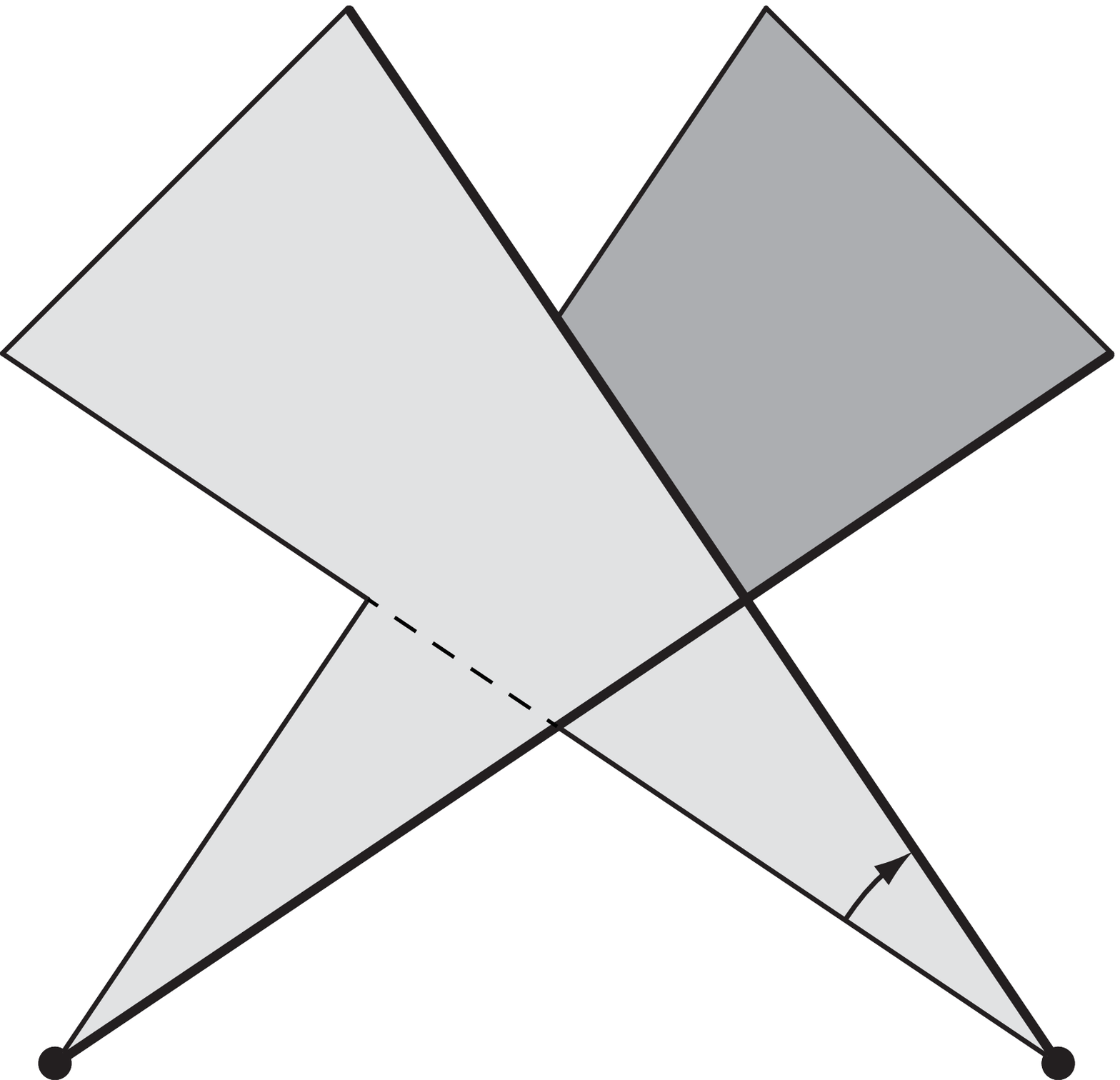}}\quad \ %\\ \vspace{0.5cm}
\subfigure[]{\includegraphics[width=0.29\linewidth]{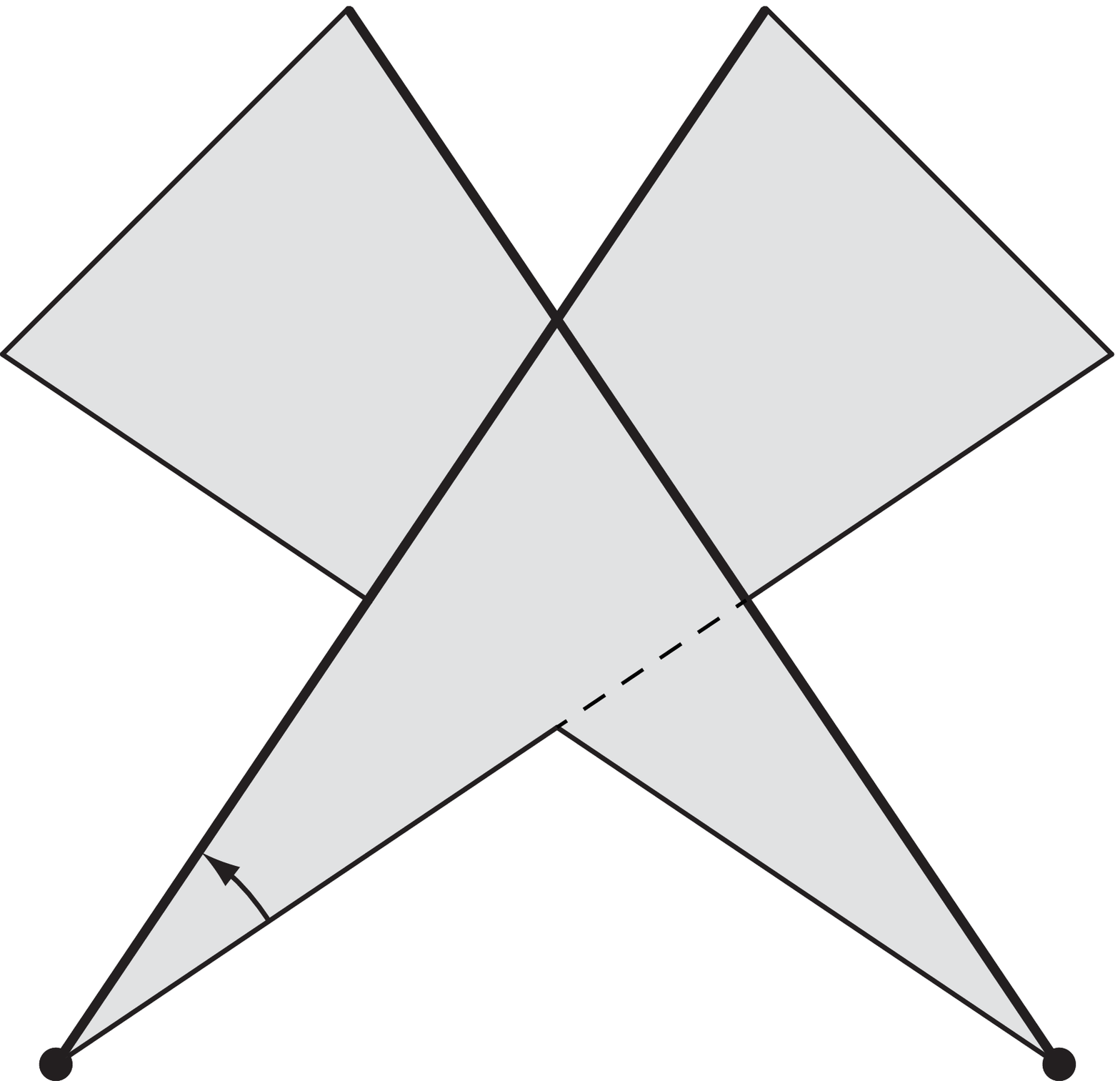}}%\qquad \qquad
%\subfigure[]{\includegraphics[width=0.35\linewidth]{fig02d.eps}}
\caption{A searchable instance of \SSP\ with no monotone search schedule. There are several search schedules (only one is depicted), but all of them rotate a searchlight back and forth.}
\label{fig:2}
\end{figure}

The problem of determining the computational complexity of \SSP\ was not directly addressed in \cite{search1}, but has acquired more interest over time, and remains open. In \cite{bullo} it is shown that \SSP\ is solvable in double exponential time, by a discretization process that reduces the space of all possible searchlight schedules to a finite graph, which is then searched exhaustively. It is also straightforward to prove that \SSP\ belongs to \PSPACE$=$\NPSPACE, because the information contained in a node can be stored efficiently, and the graph can be searched nondeterministically. It is still unknown whether \SSP\ is \NP-hard or even in \NP. Conjecture~3.1 in \cite{bullo} states that any searchable instance of \SSP\ is also searchable by rotating each searchlight either exclusively clockwise or exclusively counterclockwise, from some initial position. Should this hold true, it would imply that \SSP\ belongs at least to \NP, but we can provide simple counterexamples, such as the one showed in Figure~\ref{fig:2}: either searchlight has to sweep back and forth, in order to search the polygon.

In \cite{viglietta} the author (with M.~Monge) extended the basic model to 3-dimensional polyhedra, as opposed to polygons. The traditional point guards then become segment guards, casting half-planes (as opposed to 1-dimensional rays), which can rotate with one degree of freedom. After providing some geometric motivations for the choice of the model and pointing out some of its basic features, the authors considered the optimization problem of searching a polyhedron in the shortest time, and proved its strong \NP-hardness.

\subsubsection*{Our contribution}
In this paper we further develop the theory of searching polyhedra by rotating half-planes: we expand on the results contained in \cite{viglietta} and we prove several new theorems. In Section~\ref{model} we give a careful and thorough definition of the model, which \cite{viglietta} lacked. In Section~\ref{basic} we make preliminary observations and discuss the possibility of generalizing the main features of \SSP\ to our model. Section~\ref{heuristics} is devoted to algorithms to place guards in a given polyhedron that guarantee searchability, both for orthogonal and for general polyhedra. In Section~\ref{complexity} we show the strong \NP-hardness of deciding if a polyhedron is searchable by a given set of guards, which greatly improves on the main result of \cite{viglietta}. Indeed, being able to minimize the search time (where infinite search time means unsearchable) allows a fortiori to decide searchability. We show also that deciding if a given target area is searchable (without necessarily searching the entire polyhedron) is strongly \PSPACE-hard, even for orthogonal polyhedra. Section~\ref{conclusions} contains concluding remarks and suggestions for further research.

\section{Model definition}\label{model}
\subsubsection*{Polyhedra}
For our purposes, a \emph{polyhedron} will be the union of a finite set of closed tetrahedra (with mutually disjoint interiors) embedded in $\R^3$, whose boundary is a connected 2-manifold. As a consequence, a polyhedron is a compact topological space and its boundary is homeomorphic to a sphere or a $g$-torus. $g$ is also called the \emph{genus} of the polyhedron, and by definition it is 0 if the polyhedron is homeomorphic to a ball. Moreover, the complement of a polyhedron with respect to $\R^3$ is connected. Since a polyhedron's boundary is piecewise linear, the notion of \emph{face} of a polyhedron is well-defined as a maximal planar subset of its boundary with connected and non-empty relative interior. Thus a face is a plane polygon, possibly with holes, and possibly with some degeneracies, such as hole boundaries touching each other at a single vertex. Any vertex of a face is also considered a \emph{vertex} of the polyhedron. \emph{Edges} are defined as minimal non-degenerate straight line segments shared by two faces and connecting two vertices of the polyhedron. Since a polyhedron's boundary is an orientable 2-manifold, the relative interior of an edge lies on the boundary of exactly two faces, thus determining an internal dihedral angle (with respect to the polyhedron). A \emph{notch} is an edge whose internal dihedral angle is reflex, i.e., strictly greater than $\pi$. Hence, convex polyhedra have no notches. A polyhedron is said to be \emph{orthogonal} if each one of its edges is parallel to some axis.

\emph{Visibility} with respect to a polyhedron $\mathcal P$ is a symmetric relation between points in $\R^3$: point $x$ \emph{sees} point $y$ (equivalently, $y$ is \emph{visible} to $x$) if the straight line segment joining $x$ with $y$ lies entirely in $\mathcal P$. Recall that $\mathcal P$ is a closed set, therefore such a segment could touch $\mathcal P$'s boundary, or even lie on it. When $\mathcal P$ is understood, we can safely omit any explicit reference to it.

\subsubsection*{Searchlight scheduling}
Now we can state the problem-specific definitions. For simplicity, we consider only boundary guards, because they already yield a rich and diverse theory, unlike the situation in planar \SSP.
\begin{definition}[Guard]\label{guard}A \emph{guard} in a polyhedron $\mathcal P$ is a positive-length straight line segment without its endpoints, lying on the boundary of $\mathcal P$.\end{definition}
We exclude guard endpoints, because we don't want them to see beyond notches or non-convex vertices, as the next definitions will clarify.

A guard is said to lie \emph{over} edge $e$ if it coincides with the relative interior of $e$.
\begin{definition}[Visibility region]The \emph{visibility region} $\mathcal V(\ell)$ of a guard $\ell$ in a polyhedron $\mathcal P$ is the set of points in $\mathcal P$ that are visible to at least one point in $\ell$.\end{definition}
\begin{definition}[Searchplane]A \emph{searchplane} of a guard $\ell$ is the intersection between $\mathcal V(\ell)$ and any half-plane whose bounding line contains $\ell$.\end{definition}
Consequently, every guard has a searchplane for every possible direction of the half-plane generating it, and the union of a guard's searchplanes coincides with its visibility region.

\begin{figure}[ht]
\centering
\includegraphics[scale=0.7]{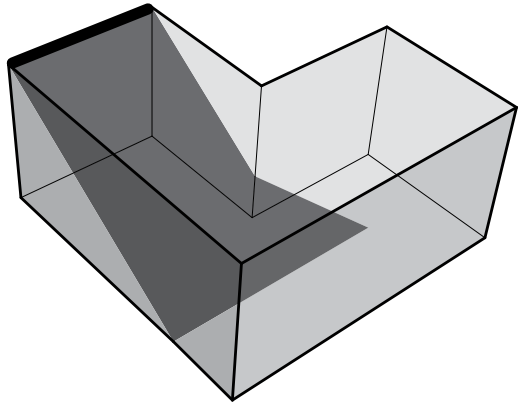}
\caption{A guard and one of its searchplanes, depicted as a thick line and a dark surface, respectively.}
\label{fig:3}
\end{figure}

If a searchplane is just a line segment, it's said to be \emph{trivial}, and the corresponding direction is said to be \emph{blind} for its guard. We arbitrarily define a \emph{left} and \emph{right} side for each guard, and we call \emph{leftmost position} the leftmost non-blind direction, for each guard. Similarly, we define the \emph{rightmost position} of every guard. Observe that the leftmost and rightmost positions are well-defined, because the polyhedron is a closed set, and every direction aiming straight at its exterior is blind for a guard, even if the endpoints of (the topological closure of) the guard lie on reflex edges or vertices. This is because we didn't include endpoints in Definition~\ref{guard}, a choice motivated also by Theorem~\ref{one}. Conversely, every other direction is not blind, since the corresponding searchplanes must contain either some internal points of the polyhedron, or some points in the relative interior of a face.

\begin{definition}[Schedule]A \emph{schedule} for a guard $\ell$ is a continuous function $f_\ell: [0,T] \to S^1$, where $T\in \R^+$ and $S^1$ is the unit circle.\end{definition}
Intuitively, $f_\ell(t)$ expresses the orientation of the guard at time $t\in [0,T]$, which is the angle at which $\ell$ is aiming its searchlight. In other words, $\ell$ is able to emit a half-plane of light in any desired direction, and to rotate it continuously about the axis defined by $\ell$ itself. We will say that, at time $t$, $\ell$ is \emph{aiming its searchlight} at point $x$ if the orientation expressed by $f_\ell(t)$ corresponds to a searchplane of $\ell$ containing $x$ (assuming that one exists).

For the following definitions, we stipulate that a polyhedron $\mathcal P$ is given, along with a finite multiset of guards, each of which is provided with a schedule.
\begin{definition}[Illuminated point]A point is \emph{illuminated} at a given time if some guard is aiming its searchlight at it.\end{definition}
\begin{definition}[Contaminated point, clear point]A point $x$ is \emph{contaminated} at time $t$ if there exists a continuous function $g: [0, t] \to \mathcal P$ such that $g(t)=x$ and there is no time $t'\in [0, t]$ at which $g(t')$ is illuminated. A point that is not contaminated is said to be \emph{clear}.\end{definition}
%Due to the compactness of $\mathcal P$ and the finiteness of the guard multiset, a maximal connected region of $\mathcal P$ without illuminated points is either all clear or all contaminated.
It follows that a maximal connected region of $\mathcal P$ without illuminated points is either all clear or all contaminated.
\begin{definition}[Search schedule]A set of schedules of the form $f_\ell: [0,T] \to S^1$, where $\ell$ ranges over a finite guard multiset in a polyhedron $\mathcal P$, is a \emph{search schedule} if every point in $\mathcal P$ is clear at time $T$.\end{definition}
\begin{sloppypar}Next we define the \textsc{3-dimensional Searchlight Scheduling Problem} (\TSSP).\end{sloppypar}
\begin{definition}[\TSSP]\emph{\TSSP}\ is the problem of deciding if a given multiset of guards in a given polyhedron has a search schedule.\end{definition}
An instance of \TSSP\ is said to be \emph{searchable} or \emph{unsearchable}, depending on the existence of a search schedule for its guards.

Occasionally in our constructions we will need two guards to be coincident, hence we consider multisets of guards, as opposed to sets.

It is obvious, from these definitions, that \TSSP\ is not easier than \SSP.
\begin{proposition}$\SSP \leqslant_\L \TSSP$.\end{proposition}
\begin{proof}Any polygon can be extruded to a prism, while each point guard can be transformed into a segment guard by stretching it parallel to the prism's sides.\end{proof}
Notice, though, that not all reasonable models of 3-dimensional guards would yield such an immediate reduction.

Since an instance of \TSSP\ is trivially unsearchable if its guards can't see the whole polyhedron, we want to exclude those instances.
\begin{definition}[Viable instance]\label{viable}An instance of \TSSP\ is \emph{viable} if every point of the polyhedron belongs to the visibility region of at least one guard.\end{definition}
We will see that there are viable but unsearchable instances of \TSSP.

Finally, a relevant role is played by a special type of guard.
\begin{definition}[Exhaustive searchplane]A searchplane of a guard in a polyhedron $\mathcal P$ is \emph{exhaustive} if it is a closed set whose relative boundary lies entirely on $\mathcal P$'s boundary.\end{definition}
Consider a guard $\ell$ with a non-trivial searchplane $S$, and let $\alpha$ be the plane containing $S$. Then $S$ is exhaustive if and only if it coincides with the connected component of $\alpha \cap \mathcal P$ containing $\ell$. Notice that the searchplane depicted in Figure~\ref{fig:3} is not exhaustive, because one of its edges lies in the interior of the polyhedron (plus, it's not a closed set: recall that guards have no endpoints). 
\begin{definition}[Exhaustive guard]A guard is \emph{exhaustive} if all its searchplanes are exhaustive.\end{definition}
Intuitively, an exhaustive guard is similar to a traditional boundary guard from \SSP\ in simple polygons, in that its searchlight provides at any time an effective barrier which cannot be crossed by the intruder just by walking past its borders. The importance of such guards in developing search algorithms will be clear shortly.

\section{Basic results}\label{basic}
\subsubsection*{Counterexamples}
The most noteworthy aspect of our guard model is that searchplanes, as opposed to searchlight rays emanating from boundary guards in \SSP, may fail to disconnect a polyhedron when aimed at its interior, regardless of its genus. As it turns out, this is the main reason why \TSSP\ seems harder than \SSP, in that exploiting such a property will enable the relatively simple \NP-hardness proof in Section~\ref{complexity}, as well as the construction of several counterexamples to positive statements about \SSP.

\begin{figure}[ht]
\centering
\includegraphics[width=0.35\linewidth]{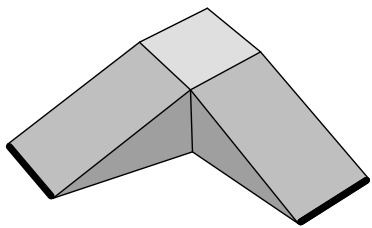}
\caption{A searchable instance of \TSSP\ with no sequential search schedule. Thick lines mark guards.}
\label{fig:4}
\end{figure}

For example, the reduction of the search space to \emph{sequential schedules} (i.e., schedules in which the guards sweep in turns) given in \cite{bullo} is no longer possible. Figure~\ref{fig:4} shows an instance of \TSSP\ whose two guards are forced to turn their searchlights simultaneously, or else they would create gaps in the illuminated surface which would result in the recontamination of the whole polyhedron.

\begin{figure}[ht]
\centering
\includegraphics[scale=0.5]{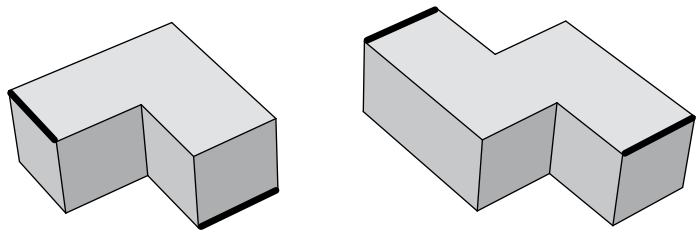}
\caption{Two unsearchable instances of \TSSP\ whose guards solve the \ART.}
\label{fig:5}
\end{figure}

Moreover, in spite of the searchability of all \SSP\ instances whose guards lie on the boundary and collectively see the whole polygon (see \cite{search1}), it is easy to construct viable but unsearchable instances of \TSSP, such as those in Figure~\ref{fig:5}. Indeed, whenever the two guards attempt to clear the center (in either of these two instances), they fail to disconnect the polyhedron, since their searchplanes are not coplanar, which results in the recontamination of the entire instance. In Section~\ref{complexity} we will provide more sophisticated unsearchable but viable instances of \TSSP.

\subsubsection*{Exhaustive guards}
\begin{figure}[ht]
\centering
\subfigure[]{\label{fig:6a}\includegraphics[scale=0.45]{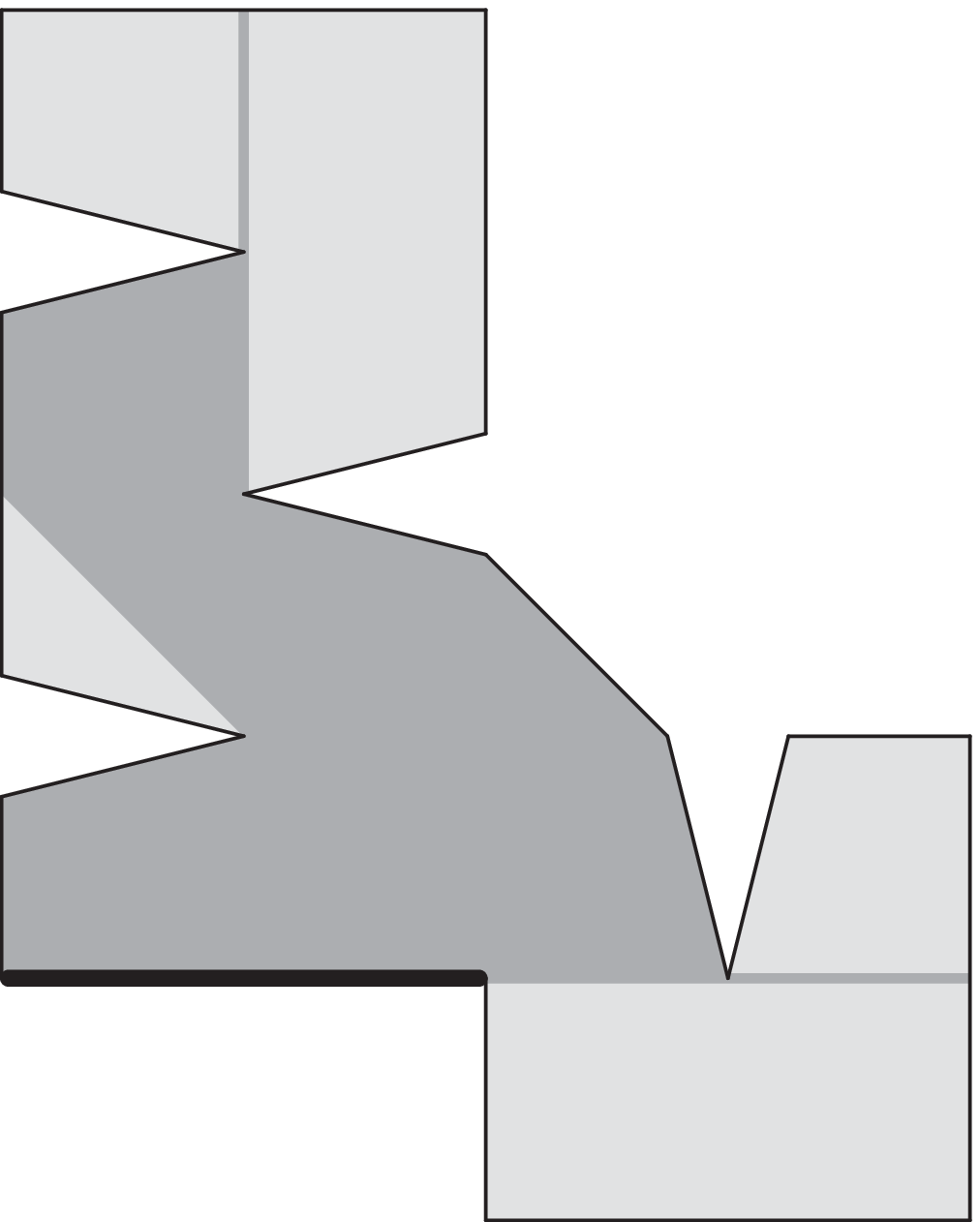}}\qquad \qquad \quad
\subfigure[]{\label{fig:6b}\includegraphics[scale=0.45]{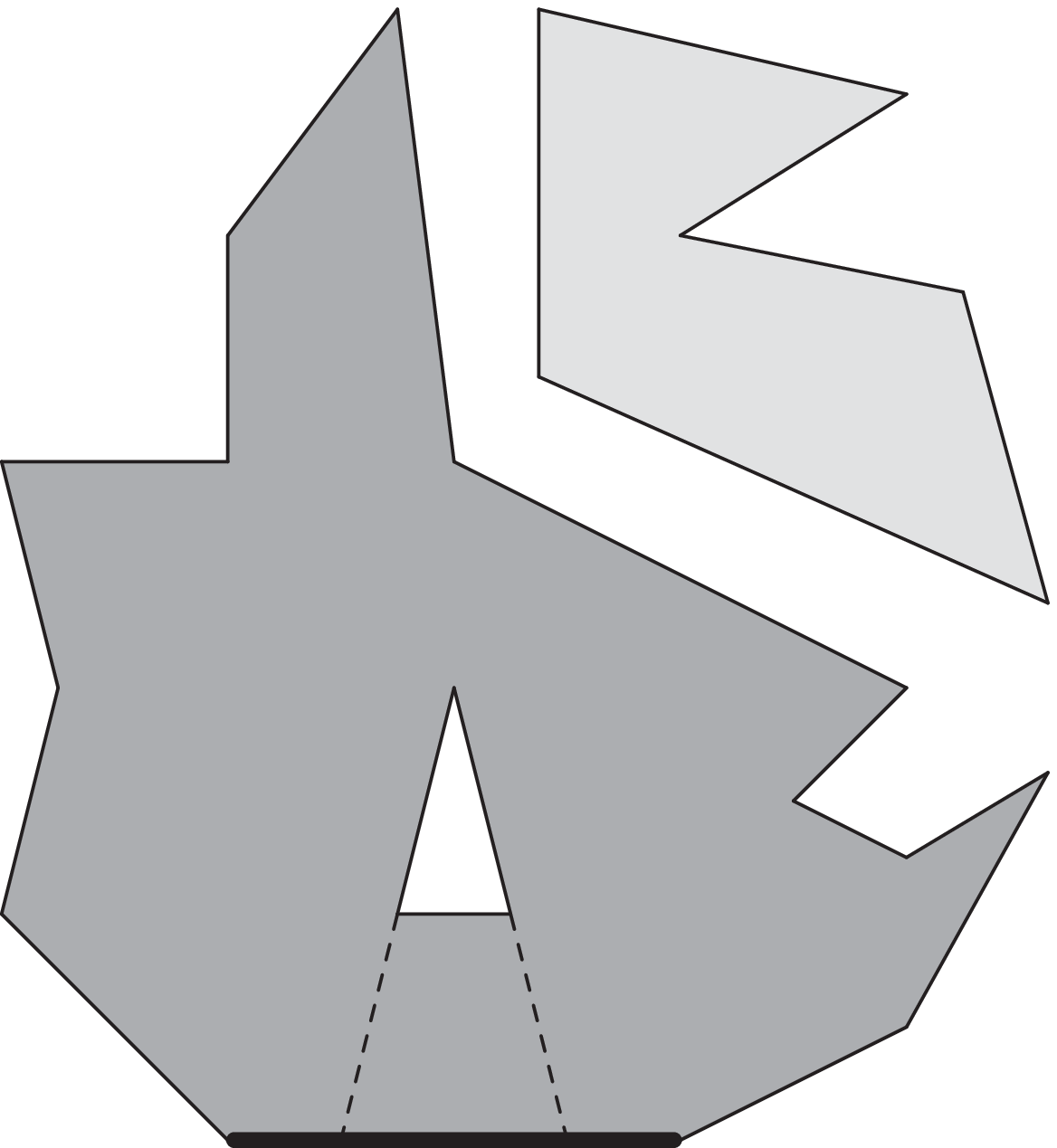}}
\caption{Two sections of polyhedra, with searchplanes represented as dark regions. The searchplane in \subref{fig:6a} has two dangling segments, while the searchplane in \subref{fig:6b} is exhaustive but not simply connected.}
\label{fig:6}
\end{figure}

Notice that all the previous counterexamples employ guards that are not exhaustive. Conversely, it comes as no surprise that employing only exhaustive guards yields positive results. To see why, let's give a characterization of the different shapes a searchplane can take with respect to the surrounding polyhedral environment. The topological closure of a searchplane is always a polygon, perhaps with holes, perhaps with some additional segments sticking out radially, and the whole searchplane is visible to some line segment lying on its external boundary, which would be the guard emanating it (refer to Figure~\ref{fig:6}). There may be intersections between a searchplane's relative interior and the polyhedral boundary, which could be collections of polygons, straight line segments, and isolated points. But what's central for our purposes is the searchplane's relative boundary, which may entirely lie on the polyhedron's boundary, or may not. If it does, and the searchplane is a closed set, then the only way an intruder could travel from one side of the searchplane to the other, without crossing the light and being caught, would be to take a detour through a suitable handle of the polyhedron. In particular, in 0-genus polyhedra, that would be impossible. In other words, any exhaustive guard aiming its searchlight at the interior of a 0-genus polyhedron, disconnects it.

On the other hand, if $\ell\subset\mathcal P$ is exhaustive, the topological closure of $\mathcal V(\ell)$ is always a polyhedron, perhaps with some dangling polygons. Of course, the boundary of $\mathcal V(\ell)$ may contain polygons that do not lie on $\mathcal P$'s boundary. But, because $\ell$ is exhaustive, every such polygon is entirely contained in some searchplane of $\ell$, and the corresponding searchlight position will be called \emph{critical} for $\ell$. Every exhaustive guard has only a finite number of critical positions.

Now, as an exhaustive guard $\ell$ in a 0-genus polyhedron starts turning from its leftmost position toward the right, every point that it illuminates will remain clear forever, unless the illuminated searchplane becomes tangent to some region of the polyhedron which is not in $\mathcal V(\ell)$ and which would be responsible for recontamination, once the tangency is crossed by the searchlight. This happens exactly when $\ell$ reaches a critical position.

\subsubsection*{One-way sweeping}
We now have the tools required to generalize the one-way sweep strategy for guards in simple polygons (see \cite{search1}) to work with exhaustive guards in simply connected polyhedra.
\begin{theorem}\label{exhaustive}Every viable 0-genus instance of \TSSP\ whose guards are exhaustive is searchable.\end{theorem}
\begin{proof}We first sketch a search schedule before detailing it further. Select any guard $\ell$ and start turning it rightward from its leftmost position. As soon as it reaches a critical position, it means that some \emph{subpolyhedron} $\mathcal R \subset \mathcal P$ has been encountered that is invisible to $\ell$. So stop turning $\ell$ and select another guard to continue the job. Proceed recursively until $\mathcal R$ is clear, then turn $\ell$ rightward again, stopping at every critical position, until the entire polyhedron is clear.

At any time, there is a clear region of $\mathcal P$ that is steadily growing, and a \emph{semiconvex subpolyhedron} $\mathcal R$ \emph{supported} by a set of guards $L$ that is being cleared by some guard not in $L$, while the guards in $L$ hold their searchlights fixed. Intuitively, the term \textquotedblleft semiconvex\textquotedblright\ is used because the only points of non-convexity of such a polyhedron lie on $\mathcal P$'s boundary. For a formal definition of \emph{semiconvex subpolygon} and \emph{support}, refer to \cite{search1} or \cite{bullo}. Extending these definitions to polyhedra is straightforward, since we're considering only exhaustive guards in 0-genus polyhedra. Thus, part of the boundary of $\mathcal R$ coincides with $\mathcal P$'s boundary, while the remaining part is determined by the searchplanes of the guards in $L$. Moreover, all the guards in $L$ are in a critical position, waiting for $\mathcal R$ (or some larger semiconvex subpolyhedron of $\mathcal P$, supported by a subset of $L$) to be cleared. It follows that none of the guards in $L$ can see any point in the interior of $\mathcal R$.

Hence, there must be some guards not in $L$ that can see an internal portion of $\mathcal R$, otherwise the problem instance wouldn't be viable. We select one of them, say $\ell'$, and start sweeping $\mathcal R$ from left to right. Notice that a searchplane bounding $\mathcal R$ could have holes, and thus $\mathcal R$ itself could have strictly positive genus. But that does not affect our invariants, because every searchplane of $\ell'$ passing through $\mathcal R$'s interior still disconnects it, or it wouldn't even disconnect $\mathcal P$. As a consequence, the points in $\mathcal R$ that are illuminated by $\ell'$ never get recontaminated as $\ell'$ continues its sweep.

Again, whenever $\ell'$ reaches a critical position, it stops there until the semiconvex subpolyhedron $\mathcal R' \subset \mathcal R$ supported by $L\cup \{\ell'\}$ has been cleared by some other guard, and so on recursively. Since every guard has only a finite number of critical positions, eventually $\mathcal P$ gets cleared.\end{proof}
Remarkably enough, the core argument supporting the planar one-way sweep strategy of \cite{search1} applies also to our polyhedral model, where there is no well-defined global concept of clockwise rotation.

\subsubsection*{Sequentiality}
In addition to this, a version of the main result of \cite{bullo} also extends to instances of \TSSP\ with exhaustive guards. We say that a search schedule is \emph{critical} and \emph{sequential} if at most one guard is turning at any given time, and guards stop or change direction only at critical positions, or at their leftmost or rightmost positions.
\begin{corollary}Every searchable 0-genus instance of \TSSP\ whose guards are exhaustive has a critical and sequential search schedule.\end{corollary}
\begin{proof}Since the instance is searchable, it must be viable. Thus, the search schedule detailed in the proof of Theorem~\ref{exhaustive} applies, which is indeed critical and sequential.\end{proof}

\subsubsection*{Searching with one guard}
As a further application of the concept of exhaustive guard, we characterize the searchable instances of \TSSP\ having just one guard, which turn out to be all the viable ones.
\begin{theorem}\label{one}Every viable instance of \TSSP\ with just one guard is searchable.\end{theorem}
\begin{proof}Since the instance is viable, the visibility region of its only guard $\ell$ coincides with the whole polyhedron $\mathcal P$. Let $\ell'$ be the maximal straight line segment contained in $\mathcal P$ and containing $\ell$. Then $\ell'$ entirely belongs to the boundary of $\mathcal P$. Indeed, if a point $x\in \ell'$ lay strictly in the interior of $\mathcal P$, then also some neighborhood of $x$ would. Recall that $\ell$ has a range of blind directions past its leftmost and rightmost position, where all its searchplanes degenerate to the single line $\ell'$. In all those directions, part of the neighborhood of $x$ would lie outside $\mathcal V(\ell)$, contradicting the viability of the instance.

\begin{figure}[ht]
\centering
\psfrag{a}{$\alpha$}
\psfrag{x}{$x$}
\psfrag{s}{$S$}
\psfrag{l}{$\ell$}
\psfrag{j}{$\ell'$}
\includegraphics[scale=0.5]{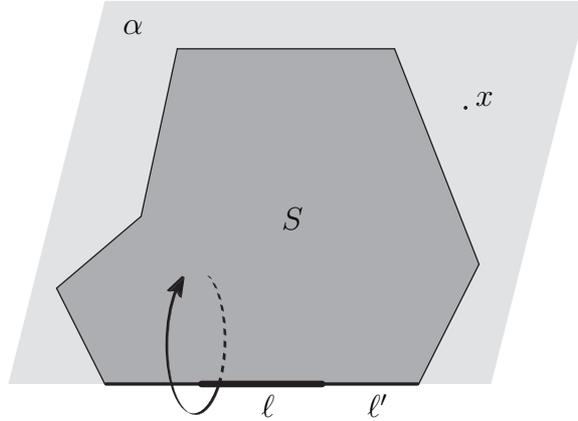}
\caption{An illustration of the proof of Theorem~\ref{one}.}
\label{fig:n1}
\end{figure}

Consider a searchplane $S$, lying on a half-plane $\alpha$ whose bounding line contains $\ell$ (refer to Figure~\ref{fig:n1}). If a point $x \in \alpha \cap \mathcal P$ lay outside $S$, then $x$ would necessarily be covered by another searchplane, because $\mathcal V(\ell)=\mathcal P$. But the only points in $\alpha$ that could lie on a searchplane different from $S$ would be those in $\ell'$, because any two searchplanes of $\ell$ intersect just on $\ell'$. Nonetheless, $\ell'$ belongs to $S$ too, which yields a contradiction. Hence $S=\alpha \cap \mathcal P$ and also $\ell'$ lies on the boundary of $\mathcal P$, implying that $\ell$ is an exhaustive guard with no critical positions. Moreover, $S$ disconnects $\mathcal P$ if it intersects its interior. Suppose by contradiction that an intruder could walk from one side of $S$ to the other side. Since $S=\alpha \cap \mathcal P$, the intruder would necessarily have to take the long route around $\ell'$ (i.e., opposite to $\alpha$, see Figure~\ref{fig:n1}), and by doing so it would cross all the half-planes in the blind directions of $\ell$. But the only points of $\mathcal P$ that belong to such half-planes are those in $\ell'$, so the intruder would be caught in any case.

It follows that turning $\ell$ from its leftmost position to its rightmost position produces a search schedule.\end{proof}
Notice that the above characterization includes polyhedra of any genus, not just 0. Also notice that, had we included endpoints in Definition~\ref{guard}, the statement of Theorem~\ref{one} would have been false. Indeed, the viability assumption would have been satisfied by more \TSSP\ instances, including unsearchable ones (such as a cube with a \textquotedblleft pyramidal well\textquotedblright\ pointing inside, and a guard over a notch).

\subsubsection*{Checking exhaustiveness}
We conclude this Section by sketching an argument supporting the claim that the conditions of Theorem~\ref{exhaustive} are practically checkable.
\begin{proposition}The exhaustiveness of a guard $\ell\subset\mathcal P$ can be decided in time polynomial in the size of $\mathcal P$.\end{proposition}
\begin{proof}First of all, the exhaustiveness of a searchplane $S$ generated by a half-plane $\alpha$ can be efficiently checked by computing the polygonal section $P=\alpha \cap \mathcal P$, and then computing the boundary of $S$ by drawing lines through every pair of vertices of $P$. Once the boundary of $S$ is known, its exhaustiveness can be checked easily.

Now, determining if $\ell$ is exhaustive can be carried out by inspecting just a polynomial number of its searchplanes. For every face $F\subset \mathcal P$ and every non-parallel edge $e\subset \mathcal P$, we call the point of intersection between the plane containing $F$ and the line containing $e$ an \emph{event}. In particular, every vertex of $\mathcal P$ is an event. Imagine turning the searchlight of $\ell$ from its leftmost position to the rightmost: it is straightforward to see that the exhaustiveness of the illuminated searchplane can change only when the searchlight crosses an event. Hence, it suffices to check the exhaustiveness of every searchplane corresponding to an event, plus one searchplane for each interval between two consecutive events. The number of searchplanes to check is thus polynomial.\end{proof}

\section{Search strategies}\label{heuristics}
In this Section we attempt to solve the problem of efficiently placing guards in a given polyhedron in order to make it searchable, while partially settling Conjecture~1 in \cite{viglietta} (i.e., that any polyhedron with a guard over each notch is searchable). Our general approach employs $r^2$ guards for polyhedra with $r>0$ notches, which we believe to be asymptotically suboptimal. However, we also prove that just $r$ guards are sufficient for orthogonal polyhedra. A related goal is to minimize the \emph{search time}, which is the total time of a search schedule, assuming that the maximum angular speed of every guard is $2\pi~\mathrm{rad} / \mathrm{sec}$ (i.e., the set of its legal schedules is restricted to Lipschitz continuous functions, whose Lipschitz constants match an angular speed of $2\pi~\mathrm{rad} / \mathrm{sec}$). We will show that occasionally it is possible to trade guards for search speed, to some extent.

\subsubsection*{Minimizing guards}
The problem of minimizing the number of guards required to search a given polyhedron is strongly \NP-hard, even for 0-genus orthogonal polyhedra. The problem is strongly \NP-hard also if the guards are required to lie over edges, or just over notches. Indeed, the same approach used in \cite{lee} to show the strong \NP-hardness of the \ART\ with vertex guards in general polygons can be employed, with some additional adjustments, to construct a simple orthogonal polygon with the same properties. By subsequently extruding the polygon into an orthogonal prism and after some minor modifications, the same hardness result can be obtained for edge guards, as well. Finally, passing from the \ART\ to \TSSP\ is almost automatic because, as proved in \cite{search1}, boundary point guards can search a simple polygon if and only if they solve the \ART.

\subsection{Searching general polyhedra}
\subsubsection*{Open edge guarding}
We first need to consider a planar \ART\ for \emph{open edge guards} (i.e., edge guards without endpoints), which has apparently not been previously explored. Specifically, we want to select some edges of a given polygon $P$, in such a way that each point of $P$ is visible to an \emph{internal} point of at least one of the selected edges. For technical reasons (see the proof of Theorem~\ref{heur}), we also want to force the selection of a specific edge.
\begin{lemma}\label{edges}Every polygon with $r$ reflex vertices, $h$ holes and a distinguished edge $e$ is guardable by at most $r-h+1$ open edge guards, one of which lies over $e$.\end{lemma}
\begin{proof}Let $P$ be a polygon, select any reflex vertex $v$ and draw the bisector of the corresponding internal angle, until it again hits the boundary of $P$. If the ray hits another vertex, slightly rotate it about $v$, so that it instead hits the interior of an edge. Two situations can occur: either $P$ gets partitioned in two parts, with $r-1$ total reflex vertices and $h$ total holes, or two connected components of the boundary are joined, so that $P$ loses both a hole and a reflex vertex.

\begin{figure}[ht]
\centering
\psfrag{e}{$e$}
\subfigure[]{\label{fig:7a}\includegraphics[scale=0.35]{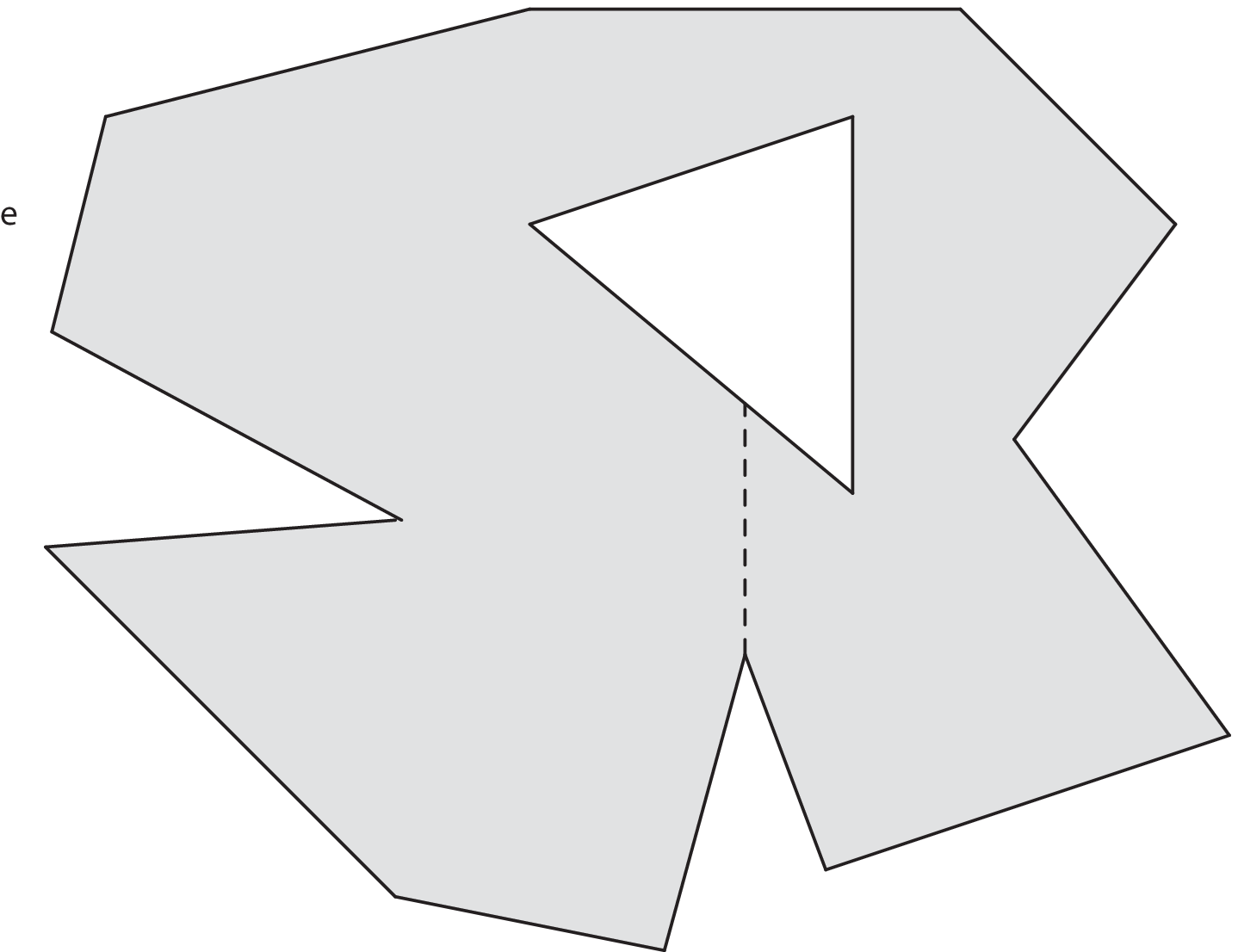}}\qquad \qquad
\subfigure[]{\label{fig:7b}\includegraphics[scale=0.35]{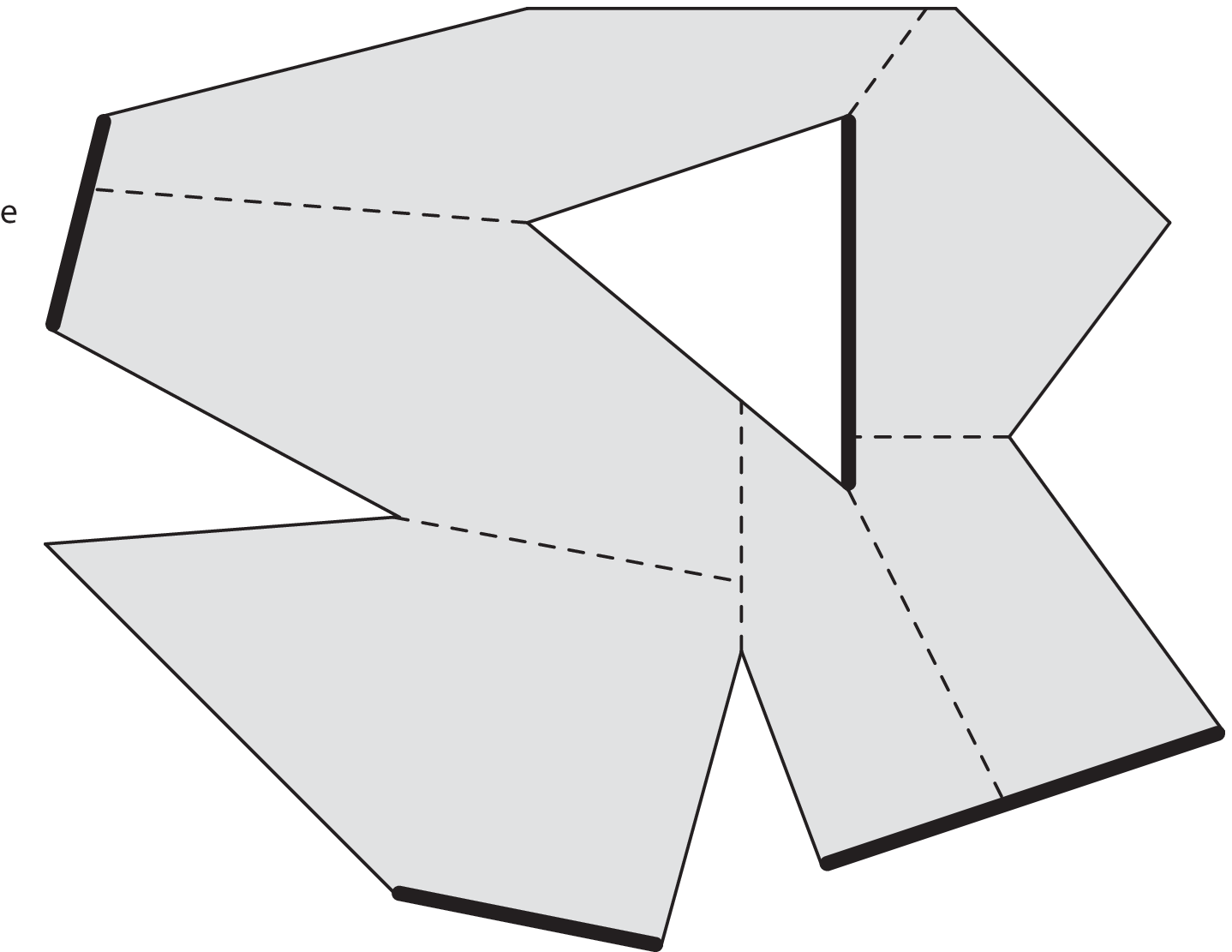}}
\caption{An example of the construction in Lemma~\ref{edges}. In \subref{fig:7a} the first step is shown, where the dotted line acts as a degenerate edge of the resulting polygon. In \subref{fig:7b} the final partition is shown, with the selected edges represented as thick lines.}
\label{fig:7}
\end{figure}

Repeat the process inductively on the resulting polygons, until no reflex vertex remains. Notice that polygonal boundaries may be \emph{degenerate} in the intermediate steps of this construction, meaning that a single segment should occasionally be regarded as two coincident segments (refer to Figure~\ref{fig:7}). $P$ is now partitioned into convex pieces, which of course have no holes. Thus, during the process, the number of holes decreased $h$ times, while the number of pieces in the partition increased $r-h$ times, resulting in $r-h+1$ convex polygons. Additionally, each polygon has at least one edge lying on $P$'s boundary, and conversely every internal point of each edge of $P$ sees at least one complete region. Place a guard on $e$, thus guarding at least one region of the partition. For each unguarded (or partially guarded) region, choose an edge of $P$ that completely sees it, and place a guard on it. As a result, $P$ is completely guarded and at most $r-h+1$ open edge guards have been placed.\end{proof}

The previous bound is asymptotically best possible, because polygons with $r$ reflex vertices can be constructed which require $r$ open edge guards, for every $r$.

\subsubsection*{Search strategy}
\begin{theorem}\label{heur}Any polyhedron with $r>0$ notches can be searched by at most $r^2$ suitably placed guards.\end{theorem}
\begin{proof}Let $\mathcal P$ be a non-convex polyhedron. We first partition it into convex regions by placing at most $r^2$ guards, then we show that some guards can be turned in a certain order to clear every piece of the partition.

\paragraph{First partition.}
Let $e$ be a notch of $\mathcal P$, and let $\alpha_e$ be a plane through $e$, close enough to its angle bisector, but not containing any vertex of $\mathcal P$ other than $e$'s endpoints. Let $\mathcal Q_e$ be the connected component of $\mathcal \alpha_e \cap \mathcal P$ containing $e$. We claim that $\mathcal Q_e$ is a polygon with at most $r-1$ reflex vertices, possibly with holes. Indeed, $e$ is an edge of $\mathcal Q_e$, and each reflex vertex of $\mathcal Q_e$ lies on a notch of $\mathcal P$. Moreover, if an endpoint of $e$ is a reflex vertex of $\mathcal Q_e$, then it belongs at least to another notch of $\mathcal P$, different from $e$. But $\alpha_e$ intersects every edge of $\mathcal P$ other than $e$ in at most one point (otherwise it would contain its endpoints, as well), hence there are at most $r-1$ reflex vertices of $\mathcal Q_e$, i.e., one for every notch of $\mathcal P$ other than $e$.

By Lemma~\ref{edges}, $\mathcal Q_e$ can be guarded by at most $r$ (2-dimensional) open edge guards, one of which lies over $e$. Equivalently, $\mathcal Q_e$ can be completely illuminated by at most $r$ suitably placed guards (with searchlights), lying on $\alpha_e$ and aimed parallel to it, one of which lies over $e$. By repeating the same construction with every other notch, at most $r^2$ guards are placed, and $\mathcal P$ is partitioned by illuminated searchplanes into convex polyhedra $\mathcal C_i$. We remark that, during our construction, previously placed searchplanes are not considered as part of $\mathcal P$'s boundary. Hence, every $\mathcal Q_{e_k}$ is indeed bounded by $\mathcal P$. This is because we need to place guards on the boundary of $\mathcal P$, and not in its interior.

(A very similar partition technique can be found in~\cite{polypart1}, under the name of \textquotedblleft naive decomposition\textquotedblright.)

\paragraph{A coarser partition.}
Consider now a slightly different partition: proceed as above by drawing angle bisectors through notches, but this time every previously drawn splitting polygon acts as a boundary. In other words, as soon as $\mathcal P$ splits into several subpolyhedra, we partition them recursively one by one, confining each split to just one subpolyhedron. So, if we consider some intermediate subpolyhedron $\mathcal P' \subseteq \mathcal P$ and select a notch $e'\subset \mathcal P'$, we look at the notch $e$ of $\mathcal P$ that contains $e'$, and let $\mathcal R_{e'}$ be the connected component of $\alpha_e \cap \mathcal P'$ (as opposed to $\alpha_e \cap \mathcal P$) containing $e'$. As a result, $\mathcal P$ is again partitioned into convex polyhedra $\mathcal D_j$, in such a way that every $\mathcal C_i$ is contained in some $\mathcal D_j$ (i.e., $\{\mathcal C_i\}$ is a \emph{refinement} of $\{\mathcal D_j\}$). Notice that, even though two different \emph{splitting polygons} $\mathcal R_{e'}$ and $\mathcal R_{e''}$ may correspond to the same notch $e$ of $\mathcal P$ (because $e$ itself has been split by a previous cut), they are nonetheless coplanar, as they both belong to the same plane $\alpha_e$.

\begin{figure}[ht]
\centering
\subfigure[$\{\mathcal C_i\}$]{\label{fig:n2a}\includegraphics[scale=0.75]{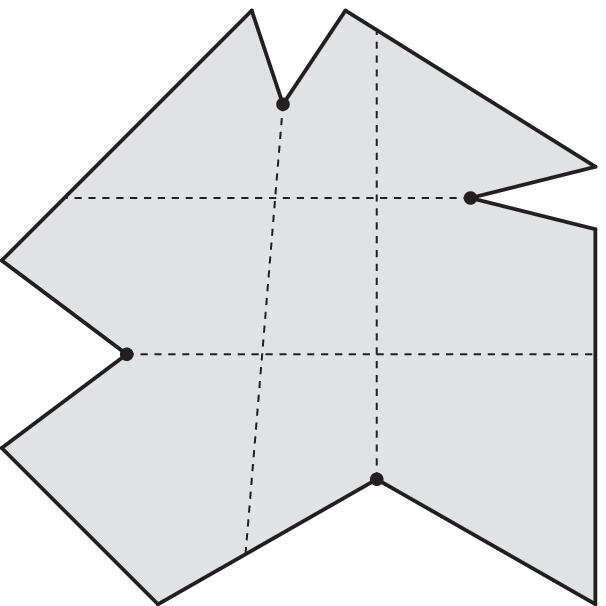}}\qquad \qquad \quad
\subfigure[$\{\mathcal D_j\}$]{\label{fig:n2b}\includegraphics[scale=0.75]{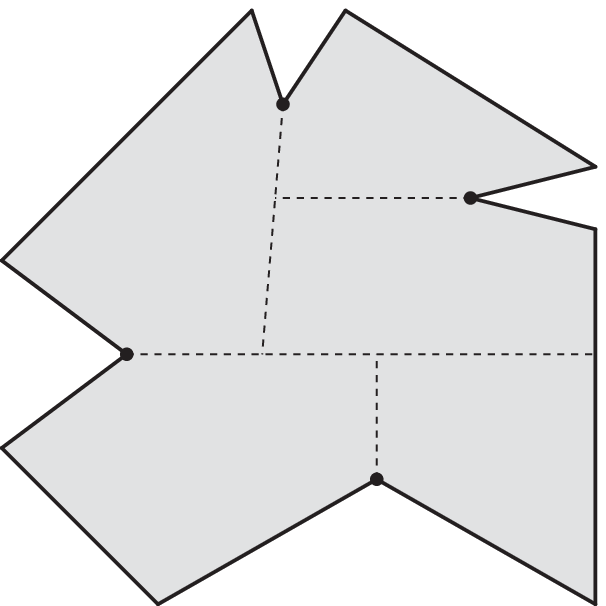}}
\caption{A comparison of the partitions $\{\mathcal C_i\}$ and $\{\mathcal D_j\}$ in a section of a polyhedron. For simplicity, only the splitting planes corresponding to visible notches are shown.}
\label{fig:n2}
\end{figure}

\paragraph{Partition tree.}
The point of having this coarser partition is that every $\mathcal D_j$ contains a subsegment of a notch of $\mathcal P$ (see Figure~\ref{fig:n2}). This property can be checked by a straightforward induction on the construction steps. More specifically, we build a tree during the splitting process, whose nodes represent the intermediate subpolyhedra, and whose arcs are marked by the splitting polygons $\mathcal R_{e}$ (see Figure~\ref{fig:n3}). Thus, the root of the tree is $\mathcal P$ and its leaves are the $\mathcal D_j$'s. Every time we draw a splitting polygon $\mathcal R_{e}$ for a subpolyhedron $\mathcal P'$ corresponding to some node $v$ of the tree, we could either decrease the genus of $\mathcal P'$, or we could partition it into two subpolyhedra $\mathcal P'_1$ and $\mathcal P'_2$, one for each side of $\mathcal R_{e}$. In the first case we just attach to $v$ an arc labeled $\mathcal R_{e}$ with a node labeled as the new polyhedron, say $\mathcal P''$. In the second case we attach to $v$ two arcs labeled $\mathcal R_{e}$, with two sibling nodes labeled $\mathcal P'_1$ and $\mathcal P'_2$. This structure is somewhat reminiscent of a Binary Space Partitioning tree. Again, like in the proof of Lemma~\ref{edges}, we have to slightly extend the notion of polyhedron, to include those with internal polygons acting as faces, resulting from non-disconnecting splits.

\begin{figure}[ht]
\centering
\psfrag{a}{$\mathcal P$}
\psfrag{b}{$\mathcal P_1$}
\psfrag{c}{$\mathcal P_2$}
\psfrag{d}{$\mathcal P_1'$}
\psfrag{e}{$\mathcal P_2'$}
\psfrag{f}{$\mathcal P_2''$}
\psfrag{g}{$\mathcal D_1$}
\psfrag{h}{$\mathcal D_2$}
\psfrag{i}{$\mathcal D_3$}
\psfrag{j}{${}_{\mathcal R_{e_1}}$}
\psfrag{k}{${}_{\mathcal R_{e_2}}$}
\psfrag{l}{${}_{\mathcal R_{e_3}}$}
\includegraphics[scale=1.1]{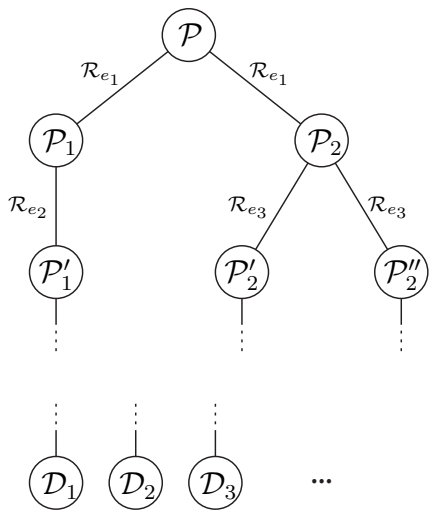}
\caption{A sketch of a partition tree.}
\label{fig:n3}
\end{figure}

\paragraph{Search schedule.}
Next we describe a search schedule for the $r^2$ guards we placed. We turn only \emph{some} of the $r$ guards that lie over the notches of $\mathcal P$, one by one, while all the other guards stay still. The order of activation of the guards is determined by the above partition tree, and the same guard could be activated more than once. The subpolyhedra of $\mathcal P$ are cleared recursively by a depth-first walk of the partition tree, starting from the root. Every time a leaf is reached through an arc labeled $\mathcal R_{e}$, its corresponding $\mathcal D_j$ is swept by the guard lying on $e$. This is feasible because $e$ lies on the boundary of $\mathcal D_j$, which is convex. After $\mathcal D_j$ is clear, the guard moves back to its initial position and the depth-first walk proceeds.

\paragraph{Correctness.}
It remains to show that no \emph{significant} recontamination can occur among the $\mathcal D_j$'s while their bounding searchlights are rotated. Suppose the depth-first walk reaches a leaf of the tree labeled $\mathcal D_j$, and let $\ell_e$ be the guard whose duty is to clear $\mathcal D_j$. Perhaps the corresponding edge $e$ of $\mathcal P$ is divided several times by the partition, so let $E$ be the set of subsegments of $e$ whose corresponding splitting planes actually appear as labels of the edges of the partition tree. Let $e'\in E$ be the subsegment of $e$ that is also an edge of $\mathcal D_j$. On the other side of $\mathcal R_{e'}$ lies a subpolyhedron $\mathcal P'$, such that $\mathcal D_j$ and $\mathcal P'$ are represented by sibling nodes in the partition tree. In order to clear $\mathcal D_j$, $\ell_e$ turns its searchlight from $\mathcal R_{e'}$ to sweep over $\mathcal D_j$, and then back to $\mathcal R_{e'}$. Since the \emph{restriction} of $\ell_e$ to $\mathcal D_j$ is exhaustive, no recontamination occurs between $\mathcal D_j$ and $\mathcal P'$ during this back-and-forth sweep  (see Figure~\ref{fig:8}).

\begin{figure}[ht]
\centering
\psfrag{e}{$\ell_e$}
\psfrag{r}{${}_{\mathcal R_{e'}}$}
\psfrag{d}{$\mathcal D_j$}
\psfrag{p}{$\mathcal P'$}
\includegraphics[scale=0.6]{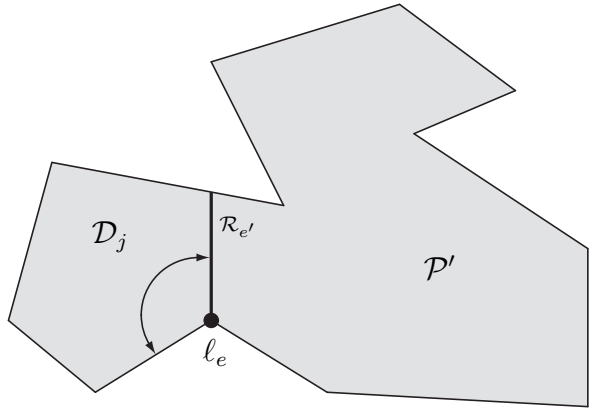}
\caption{$\mathcal P'$ is not recontaminated while $\ell_e$ sweeps $\mathcal D_j$.}
\label{fig:8}
\end{figure}

\begin{figure}[ht]
\centering
\psfrag{a}{$\alpha_e$}
\psfrag{b}{$\mathcal R_{e_1}$}
\psfrag{c}{$\mathcal R_{e_2}$}
\psfrag{d}{$\mathcal R_{e'}$}
\psfrag{e}{$e$}
\psfrag{f}{$e_1$}
\psfrag{g}{$e_2$}
\psfrag{h}{$e'$}
\psfrag{p}{$\mathcal P_1$}
\psfrag{q}{$\mathcal P_2$}
\psfrag{r}{$\mathcal P'$}
\psfrag{s}{$\mathcal D_j$}
%\psfrag{r}{${}_{\mathcal R_{e'}}$}
%\psfrag{d}{$\mathcal D_j$}
%\psfrag{p}{$\mathcal P'$}
\includegraphics[scale=0.9]{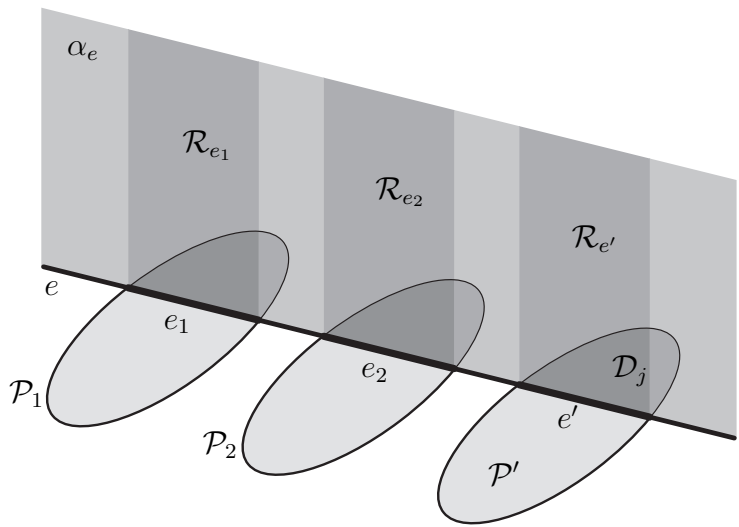}
\caption{A sketch of a splitting plane.}
\label{fig:n4}
\end{figure}

Nonetheless, recontamination could still occur between other subpolyhedra bounded by $\alpha_e$. Let $\{e_1, e_2\}\subseteq E$, and let the subpolyhedra $\mathcal P_1$ and $\mathcal P_2$ be partitioned by $\mathcal R_{e_1}$, and $\mathcal R_{e_2}$, respectively (observe that the relative interiors of $e_1$ and $e_2$ must be disjoint, by construction). Obviously, no recontamination between $\mathcal P_1$ and $\mathcal P_2$ is possible while $\ell_e$ sweeps, because $\alpha_e$ is not a common boundary (refer to Figure~\ref{fig:n4}). However, recontamination could occur within $\mathcal P_1$ (or within $\mathcal P_2$), provided that $e_1\neq e'$. As it turns out, this type of recontamination is irrelevant, because it would imply that the node labeled $\mathcal P_1$ (call it $p$) has not been reached yet by the depth-first walk in the partition tree. We set up an induction argument on the partition tree, assuming that all the subpolyhedra corresponding to previously visited leaves are still clear before $\ell_e$ sweeps. By construction, $p$ cannot be an ancestor of the node labeled $\mathcal D_j$, otherwise $\mathcal D_j$ would be a subpolyhedron of $\mathcal P_1$, while in fact their interiors are disjoint. So, had the depth-first walk reached $p$, it would have also visited its entire dangling subtree and, by inductive assumption, $\mathcal P_1$ would still be all clear. Moreover, since $\mathcal P_1$ is not bounded by $\alpha_e$, it cannot get recontaminated while $\ell_e$ sweeps. On the other hand, if the walk has not reached $p$ yet, then perhaps some portions of $\mathcal P_1$ have been accidentally cleared, but can safely be recontaminated, because the systematic clearing process of $\mathcal P_1$ has yet to start.\end{proof}

\subsubsection*{Improving search time}
The search time of our schedule could be quadratic in $r$, but we can trivially achieve constant search time by placing $r$ additional guards.
\begin{corollary}Any polyhedron with $r>0$ notches can be searched in less than 1 second by at most $r^2+r$ suitably placed guards.\end{corollary}
\begin{proof}Place $r^2$ guards as in the proof of Theorem~\ref{heur} and never move them, so that the partition $\{\mathcal D_j\}$ is always preserved. Then add a guard over every notch, and turn these additional guards simultaneously, from their leftmost position to the rightmost. Every guard turns by less than $2\pi~\mathrm{rad}$, so the search time is less than 1 second.\end{proof}

\subsection{Searching orthogonal polyhedra}
The previous result can be greatly improved if the polyhedron is orthogonal.

We define the \emph{vertical} direction (or \emph{up-down} direction) as the direction parallel to the $z$-axis. Accordingly, any direction parallel to the $xy$-plane is called \emph{horizontal}, and the direction parallel to the $x$-axis (resp.~$y$-axis) is called \emph{left-right} (resp.~\emph{front-back}) direction. Thus, the edges and faces of any orthogonal polyhedron are either vertical or horizontal, and the $x$-orthogonal (resp.~$y$-orthogonal) faces are called \emph{lateral faces} (resp.~\emph{front faces}).

\subsubsection*{Erecting fences}
\begin{figure}[ht]
\centering
\psfrag{l}{$\ell$}
\psfrag{f}{$F$}
\psfrag{g}{$F'$}
\psfrag{r}{$r$}
\psfrag{x}{$x$}
\psfrag{c}{$\mathcal C$}
\psfrag{d}{$\mathcal C'$}
\includegraphics[scale=1]{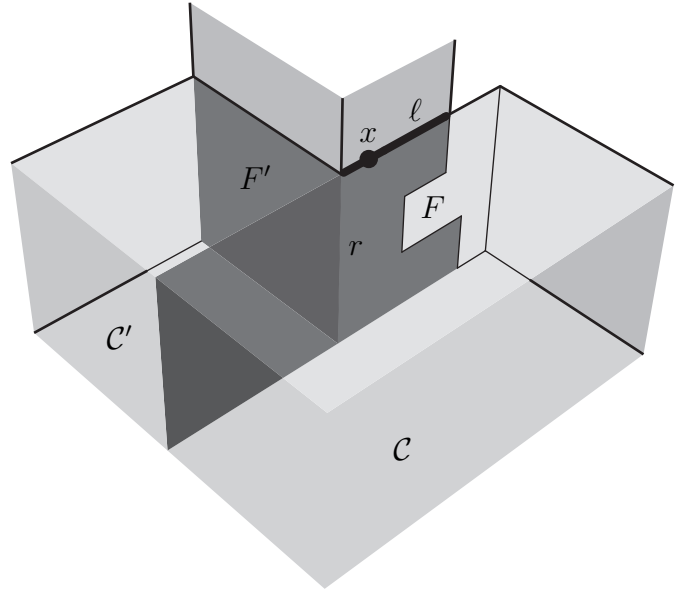}
\caption{A fence generated by $r$ during Step~\ref{step3}. Fences are shown as darker regions. The nomenclature comes from Lemmas~\ref{fences} and~\ref{cuboids}.}
\label{fig:9}
\end{figure}

We first construct a partition of a given orthogonal polyhedron $\mathcal P$ into \emph{cuboids} (i.e., rectangular boxes) by a 3-step process.
\begin{enumerate}
\item\label{step1} Each horizontal notch $r$ of $\mathcal P$ has a horizontal adjacent face and a vertical adjacent face $F$, going upwards or downwards. From every point on $r$, send a vertical ray in the direction opposite to $F$, until it again reaches the boundary of $\mathcal P$. Repeat the process with every horizontal notch, so that each generates a vertical \emph{fence}, either upwards or downwards. It's easy to see that $\mathcal P$ is partitioned by fences into orthogonal \emph{prisms} (i.e., extruded polygons) with horizontal bases.
\item\label{step2} Consider a vertical notch $r$ of $\mathcal P$ with an internal point lying on a lateral face of a prism $\mathcal Q$ generated in Step~\ref{step1}. By construction, $r$ must lie entirely on the boundary of $\mathcal Q$. Extend $r$ to a maximal segment $r'$ contained in the boundary of $\mathcal Q$. If possible, send a horizontal ray from every point of $r'$, going through the interior of $\mathcal Q$ in the left-right direction, until it reaches $\mathcal Q$'s boundary, as shown in Figure~\ref{fig:n5}. Do the same with every vertical notch of $\mathcal P$ lying on a lateral face of some prism, so that the initial partition gets further refined by these new fences. Clearly, to every such notch corresponds just one fence, which in turn goes through the interior of just one prism.

\begin{figure}[ht]
\centering
\psfrag{r}{$r$}
\psfrag{s}{$r'$}
\psfrag{q}{$\mathcal Q$}
\includegraphics[scale=0.7]{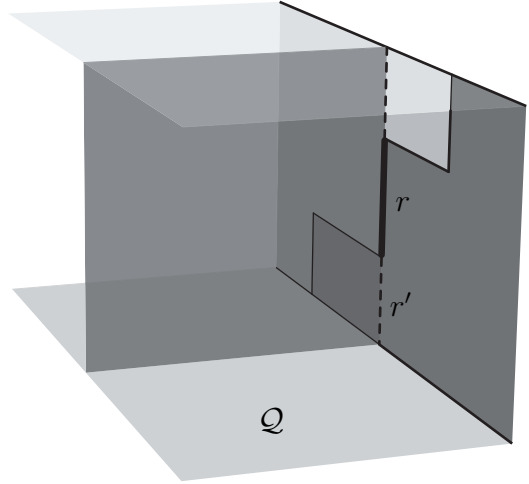}
\caption{A fence generated by $r$ during Step~\ref{step2}. Fences are shown as darker regions.}
\label{fig:n5}
\end{figure}

\item\label{step3} The pieces resulting from Step~\ref{step2} are once again prisms, perhaps with some additional non-disconnecting vertical fences. By construction, each reflex edge of such a prism is vertical, with no edges of $\mathcal P$ lying on it. Repeat the procedure in Step~\ref{step2} also with these reflex edges, sending horizontal rays and thus building vertical fences which extend laterally the front faces of the prisms, as shown in Figure~\ref{fig:9}. As a result, $\mathcal P$ gets partitioned into cuboids.
\end{enumerate}

(The term \textquotedblleft fence\textquotedblright\ is borrowed from~\cite{polypart2}).

\subsubsection*{Placing guards}
Now place a guard over every notch of $\mathcal P$. Aim horizontal guards vertically and aim vertical guards in the left-right direction, in such a way that every guard aims at the interior of $\mathcal P$. We first want to prove that the fences in our 3-step construction are \emph{coherent} with searchlights.
\begin{lemma}\label{fences}Every fence is contained in some illuminated searchplane.\end{lemma}
\begin{proof}The claim is obvious for fences constructed in Step~\ref{step1} and Step~\ref{step2}. As for fences constructed in Step~\ref{step3}, consider an edge $r$ generating one of them (see Figure~\ref{fig:9}). Recall that $r$ is a reflex edge of a prism, let it be $\mathcal Q$, in the partition obtained in Step~\ref{step1}. By the construction in Step~\ref{step2}, $r$ contains no notch of $\mathcal P$. Hence its adjacent front face $F\subset \mathcal Q$ has no intersection with the boundary of $\mathcal P$, at least in a neighborhood of $r$. But the bases of $\mathcal Q$ belong to the boundary of $\mathcal P$, because no horizontal fences were constructed. Then, at least a subsegment of a horizontal edge of $F$, sharing an endpoint with $r$, belongs to a notch of $\mathcal P$, and hence belongs to a guard $\ell$. As a consequence, $\ell$ illuminates the whole fence generated by $r$ in Step~\ref{step3}.\end{proof}
Thus, every fence \emph{belongs} to one guard. We could also incorporate each fence built in Step~\ref{step3} into its adjacent vertical fence built in Step~\ref{step1}, which belongs to the same guard. As a result, every guard \emph{generates} at most one fence.
\begin{lemma}\label{cuboids}Every cuboid in the partition belongs entirely to the visibility region of any guard whose fence bounds it.\end{lemma}
\begin{proof}If a fence built in Step~\ref{step1} bounds a cuboid, then it belongs to a guard $\ell$ located, at least partially, on its border. Indeed, fences are vertical and the upper and lower faces of a cuboid belong to faces of $\mathcal P$, and therefore $\ell$ cannot lie entirely outside the cuboid. It follows that $\ell$ sees the whole cuboid.

Fences built in Step~\ref{step2} bound exactly two cuboids each, because the fences built in Step~\ref{step3} are all parallel to them. Again, any guard generating one such fence belongs to a common edge of the cuboids that it bounds, which of course are entirely visible to the guard. 

Also fences built in Step~\ref{step3} bound exactly two cuboids each. With the same notation as in the proof of Lemma~\ref{fences}, the fence generated by $r$ bounds a cuboid $\mathcal C$ that is also bounded by $F$, and therefore contains, at least partially, guard $\ell$ on one of the horizontal edges of $\mathcal C$. $\ell$ also shares one endpoint with $r$. Moreover, the other cuboid $\mathcal C'$ is bounded also by a lateral face $F'\subset \mathcal Q$ adjacent to $r$. The interior of $F'$ has no intersection with the boundary of $\mathcal P$. Indeed, if the two had any intersection, then there would be also a vertical notch of $\mathcal P$ lying in the interior of $F'$, or on $r$. But this disagrees with the construction in Step~\ref{step2}, which eliminates all such notches. Let $x$ be a point in $\ell$ whose distance to $r$ is lower than the minimum positive difference between any two coordinates of vertices of $\mathcal P$. $x$ exists due to the finiteness of $\mathcal P$'s vertices. Of course, $x$ completely sees $\mathcal C$, because it lies on its boundary. But $x$ also sees every point in $\mathcal C'$, through the fence $F'$ (see Figure~\ref{fig:9}).\end{proof}

\subsubsection*{Search strategy}
\begin{theorem}\label{orth}Any non-convex orthogonal polyhedron with a guard over every notch is searchable.\end{theorem}
\begin{proof}Aim the guards as described above, in such a way that every fence is illuminated by some guard, by Lemma~\ref{fences}. Clearly, illuminated searchplanes induce the same partition of $\mathcal P$ into cuboids (perhaps even a finer partition). Now pick a guard $\ell$ generating a fence $F$, and let $\mathcal Q$ be the union of the cuboids bounded by $F$. Since $F$ is connected and has cuboids on both sides, it follows that $\mathcal Q$ is connected as well, and therefore it is a polyhedron. Moreover, by Lemma~\ref{cuboids}, $\mathcal Q$ entirely belongs to $\mathcal V(\ell)$. Hence, by Theorem~\ref{one}, $\mathcal Q$ can be cleared by $\ell$ while all the other guards keep their searchlights fixed. Turn $\ell$ to clear $\mathcal Q$, put $\ell$ back in its original position, and repeat the procedure for all the other guards, one at a time. Notice that every turning guard clears all the cuboids that it bounds, while the other cuboids cannot be recontaminated, because their boundaries remain fixed. Since $\mathcal P$ is both connected and non-convex, every cuboid is bounded by at least a fence, which in turn is generated by some guard. Thus, after the last guard has finished sweeping, $\mathcal P$ is completely clear.\end{proof}

\subsubsection*{Improving search time}
The search time is linear in the number of notches of $\mathcal P$, but once again we can achieve constant search time by doubling the number of guards.
\begin{corollary}Any non-convex orthogonal polyhedron with two guards over every notch is searchable in 0.75 seconds.\end{corollary}
\begin{proof}Half of the guards are positioned as in the proof of Theorem~\ref{orth} and never move, thus preserving the partition into cuboids. The other guards simultaneously sweep their visibility region from the leftmost position to the rightmost. Since every guard lies on a notch, they all have to turn by an angle of $3/2~\pi~\mathrm{rad}$, which can be done in 0.75 seconds.\end{proof}

\section{Complexity of searchability}\label{complexity}
In this Section we give two complexity theoretic results. First we show that deciding if a polyhedron is searchable by a given multiset of guards is strongly \NP-hard. Then we turn to the problem of searching only some specific parts of a polyhedron, and we prove that deciding if a given \emph{target area} is clearable at all (no matter if the rest of the polyhedron remains contaminated) is strongly \PSPACE-hard, even for orthogonal polyhedra.

\subsection{\NP-hardness of searchability}\label{nphard}
Next we prove that \TSSP\ is strongly \NP-hard by a reduction from \TSAT, thus converting a formula $\varphi$ in 3-conjunctive normal form into an instance of \TSSP\ that is searchable if and only if $\varphi$ is satisfiable.

\subsubsection*{Building blocks}
A \emph{variable gadget} is a cuboid with a guard over one edge. The two faces adjacent to the guard are called \emph{A-side} and \emph{B-side}, respectively. The guard itself is called \emph{variable guard}.

\begin{figure}[ht]
\centering
%\psfrag{a}{A}
%\psfrag{b}{B}
\subfigure[variable gadget]{\label{fig:10a}\includegraphics[scale=0.8]{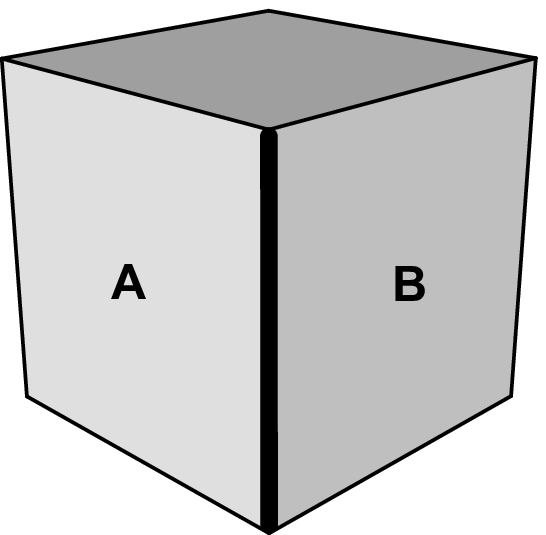}}\qquad \qquad
\subfigure[link]{\label{fig:10b}\includegraphics[scale=0.8]{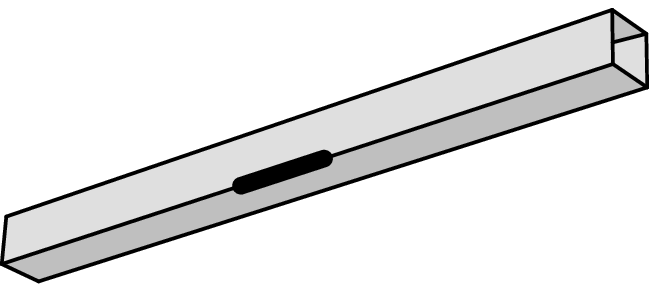}}
\caption{Two building blocks of the reduction.}
\label{fig:10}
\end{figure}

\begin{figure}[ht]
\centering
%\psfrag{a}{A}
%\psfrag{b}{B}
%\psfrag{c}{${}_{\textrm{C}}$}
\includegraphics[scale=0.8]{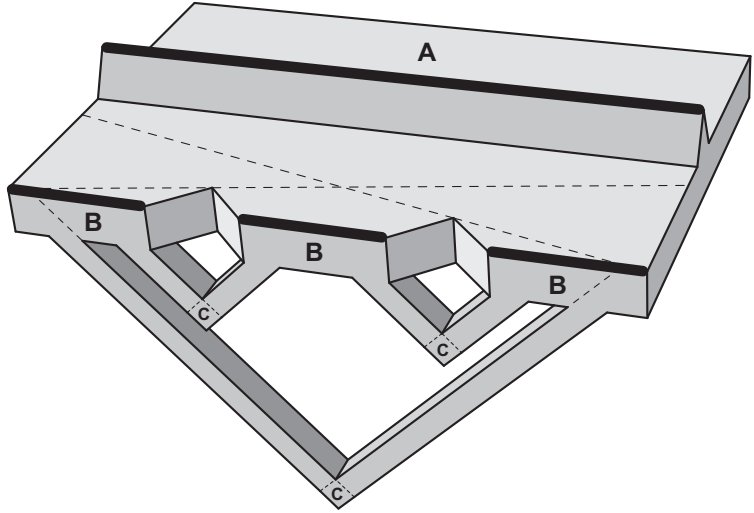}
\caption{A clause gadget.}
\label{fig:11}
\end{figure}

A \emph{clause gadget} is made of several parts, shown in Figure~\ref{fig:11}. The first part is a prism, shaped like a wide cuboid with 3 \emph{cavities} on one side. On its top there is a \emph{nook} shaped as a triangular prism lying on a side face, with a guard over the upper edge. The guard is called \emph{separator}, because its searchplanes partition the clause gadget in two regions. One of the two regions contains none of the 3 cavities, and its top face is called \emph{A-side}. On the other hand, the back face of each cavity is a \emph{B-side}, and a \emph{literal guard} lies over the top edge of each B-side. When any literal guard is aiming at the A-side of its clause gadget, it also completely closes the nook containing the separator. We define the leftmost position of the separator to be the one that is closer to the A-side. All 3 cavities are then pairwise connected by V-shaped prisms with vertical bases. When two literal guards from the same clause gadget are both aiming at their B-sides, their searchplanes intersect in an area around the bottom of the V: such area is called \emph{C-side}. Thus, every clause gadget has 3 B-sides and 3 C-sides, all coplanar.

The A-sides, B-sides and C-sides of all the gadgets are collectively referred as \emph{distinguished sides}.

To connect together all the different gadgets we use structures called \emph{links}. A link is a very thin prism with its two bases removed, and with a short \emph{link guard} lying in the middle of an edge (see Figure~\ref{fig:10}). When we wish to connect two gadgets, we cut a hole in their surfaces, and we place a link stretching from one hole to the other. We will also make sure that no guard, other than its link guard, can see inside a link. Conversely, we will arrange the links in such a way that every link guard's searchlight won't interfere with the gadget guards. So, in every gadget there will be a thin illuminated polygon jutting from each of its links, which will be easily avoidable by the intruder.  As a consequence, a link can be cleared by its guard only while both its bases are \emph{capped} by some guards lying in the adjacent gadgets.

Given a boolean formula $\varphi$ in 3-conjunctive normal form, we construct a row of variable gadgets, one for every variable of $\varphi$, and a row of clause gadgets, one for every clause of $\varphi$. We arrange the variable gadgets so that all the A-sides are coplanar, and all the B-sides are coplanar. We arrange the clause gadgets similarly, and we place the two rows of gadgets in such a way that every distinguished side of every variable gadget can see every distinguished side of every clause gadget. We also associate the $i$-th B-side of a clause gadget to the $i$-th literal in the corresponding clause of $\varphi$, for $1 \leqslant i \leqslant 3$.

Finally we add a \emph{bridge}, which is constructed like a variable gadget, but it is not associated to any variable of $\varphi$. The bridge is shaped as a long, thin pole, whose guard lies over one of the long edges, and it is arranged in such a way that its distinguished sides can see the B-side of every clause gadget.

\subsubsection*{Connections}
\begin{figure}[ht]
\centering
%\psfrag{a}{A}
%\psfrag{b}{B}
%\psfrag{c}{C}
%\psfrag{h}{variables}
%\psfrag{i}{clauses}
%\psfrag{j}{bridge}
\includegraphics[scale=0.6]{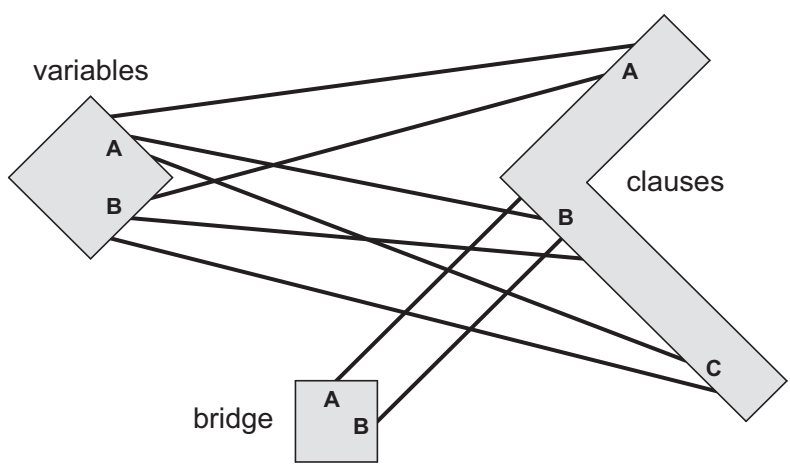}
\caption{The relative positions of the gadgets and the bridge, with their links.}
\label{fig:13}
\end{figure}

\begin{figure}[ht]
\centering
%\psfrag{a}{A}
%\psfrag{b}{B}
\subfigure[clause gadget, top view]{\label{fig:12a}\includegraphics[scale=0.7]{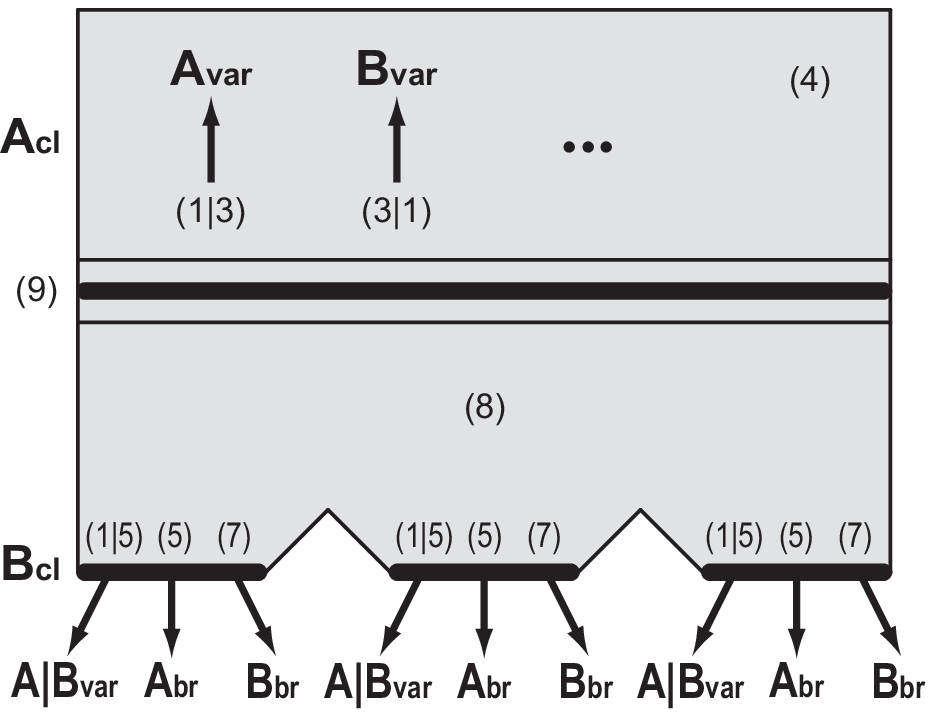}}\qquad
\subfigure[clause gadget, C-side]{\label{fig:12b}\includegraphics[scale=0.7]{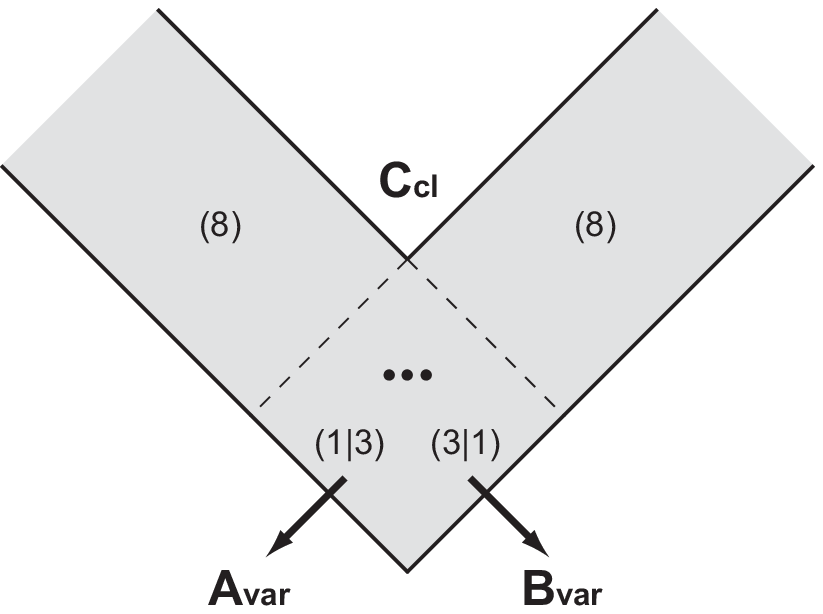}}\vspace{0.5cm}
\subfigure[variable gadget]{\label{fig:12c}\includegraphics[scale=0.7]{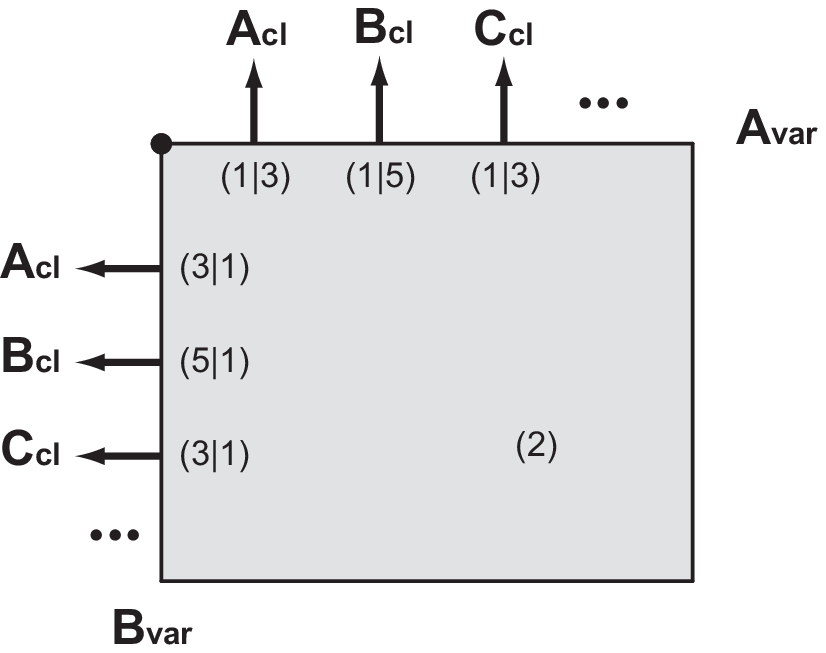}}\qquad \qquad
\subfigure[bridge]{\label{fig:12d}\includegraphics[scale=0.7]{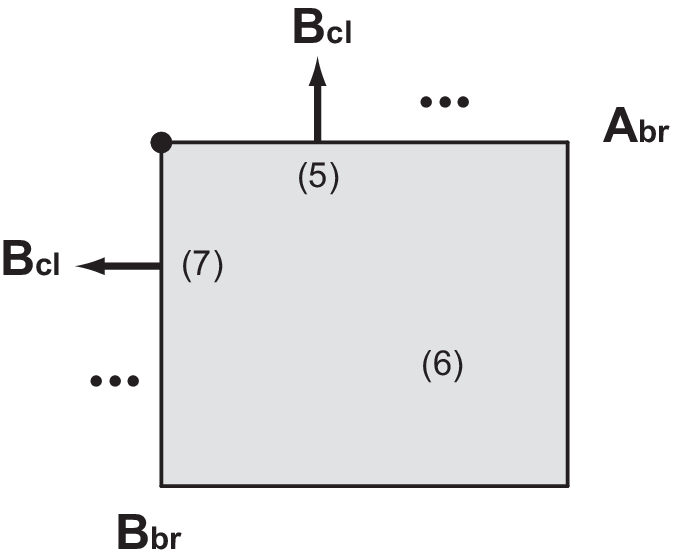}}
\caption{The connections among gadgets, and their clearing order given in Theorem~\ref{hard}. The abbreviation \textquotedblleft $\mathrm{A_{cl}}$\textquotedblright\ stands for \textquotedblleft A-side of a clause gadget\textquotedblright, etc..}
\label{fig:12}
\end{figure}

Then we connect the distinguished sides of our gadgets by placing links, as follows (refer to Figures~\ref{fig:13} and~\ref{fig:12}).
\begin{itemize}
\item Connect the A-side of every clause gadget to both the A-side and the B-side of every variable gadget.
\item Connect all the B-sides of every clause gadget to both the A-side and the B-side of the bridge.
\item Connect all the C-sides of every clause gadget to both the A-side and the B-side of every variable gadget.
\item Connect each B-side of each clause gadget to the A-side (resp.~B-side) of the variable gadget corresponding to its associated literal, if that literal is negative (resp.~positive) in $\varphi$.
\end{itemize}

We can easily position the bridge so that it's not accidentally hit by any link running between a variable gadget and a clause gadget, such as in Figure~\ref{fig:13}. We also want links to be pairwise disjoint. To achieve this, we consider any pair of intersecting links, and shrink them while translating them slightly, until their intersection vanishes. This can be accomplished without creating new intersections with other links, for example by making sure that the \textquotedblleft new version\textquotedblright\ of each link is always strictly contained in its \textquotedblleft previous version\textquotedblright.

\subsubsection*{Reduction}
\begin{theorem}\label{hard}\TSSP\ is strongly \NP-hard.\end{theorem}
\begin{proof}Given an instance $\varphi$ of \TSAT, we construct the instance of \TSSP\ described above. It is indeed a polyhedron, because the bridge and all the variable gadgets are connected to all the clause gadgets. Moreover, the number of links is quadratic in the size of $\varphi$, and we may also assume that the coordinates of every vertex are rationals, with a number of digits that is polynomial in the size of $\varphi$. Removing the intersections between links takes polynomial time as well, hence the whole construction is computable in \P.

\paragraph{Positive instances.}
If $\varphi$ is satisfiable, we choose a satisfying assignment for its variables and give a search schedule that clears our polyhedron. Initially we aim each variable guard at its A-side (resp.~B-side) if the corresponding variable is true (resp.~false) in the chosen assignment. By assumption there is at least a true literal in every clause of $\varphi$. For each clause, we pick exactly one true literal, and aim the corresponding literal guard at the A-side of its clause gadget. We aim all the other literal guards at their respective B-sides. As a result, the A-side and the 3 C-sides of every clause gadget have the corresponding links capped by the literal guards. Finally, we aim the bridge guard at its A-side and put every separator in its leftmost position.

From this starting position, we specify a search schedule in 9 steps (refer to Figure~\ref{fig:12}).
\begin{enumerate}
\item\label{stepp1} Clear all the links that are capped by some variable guard. This is possible because, by construction, the other end of every such link is capped by some literal guard as well.
\item\label{stepp2} Clear every variable gadget by turning its guard. While this happens, the literal guards retain caps on their own side of the links cleared during Step~\ref{stepp1}, thus preventing recontamination.
\item Clear the remaining links connected to the A-side or to a C-side of a clause gadget. This is now possible because all the variable guards switched side in Step~\ref{stepp2}.
\item Aim at its B-side every literal guard that is currently aiming at its A-side. One half of each clause gadget gets cleared as a result, while the separator prevents the still uncapped links on the B-side from recontaminating the clear links on the A-side.
\item Clear the remaining links connected to a variable gadget, and clear the links capped by the bridge guard.
\item Clear the bridge by turning its guard to the B-side.
\item Clear the remaining links that connect the bridge with the clause gadgets.
\item Turn all the literal guards simultaneously, thus clearing the last half of each clause gadget, and capping the upper nooks. Since the three literal guards of a clause gadget are collinear, when they move together they act as a single exhaustive guard.
\item Clear every nook by turning the separators.
\end{enumerate}
When this is done, the whole polyhedron is clear, which proves that the instance of \TSSP\ is searchable.

\paragraph{Negative instances.}
Conversely, assuming that $\varphi$ is not satisfiable, we claim that the variable gadgets can never be all simultaneously clear, no matter what the guards do.

Recall that every A-side of every clause gadget is linked to both sides of every variable gadget. Hence, as soon as the A-side of any clause gadget is not covered by at least one literal guard, all the variable gadgets get immediately recontaminated, unless they were all clear in the first place. For the same reason, no variable gadget can ever be cleared while the A-side of some clause gadget is uncovered. Similarly, if a C-side of any clause gadget is not covered by at least one literal guard, all variable gadgets get recontaminated, and none of them can be cleared.

It follows that, in order for a schedule to start clearing any variable gadget, it must ensure that each clause gadget has exactly one literal guard covering the A-side and exactly two literal guards covering the C-sides. Moreover, the literal guards that cover the A-sides must be chosen once and for all. Indeed, whenever a schedule attempts to \textquotedblleft switch gears\textquotedblright\ in some clause gadget and cover the A-side with a different literal guard, all the variable gadgets become immediately contaminated, and the search must start over.

Suppose that a schedule selects exactly one literal guard for each clause gadget, to cover its A-side. Since $\varphi$ is not satisfiable, there exist two selected literal guards $\ell_1$ and $\ell_2$ whose corresponding literals in $\varphi$ are a positive occurrence and a negative occurrence of the same variable $x$. Otherwise, if all selected literals were \emph{coherent}, setting them to true would yield a satisfying assignment for the variables of $\varphi$, which is a contradiction.

But in this case, it turns out that the variable gadget corresponding to variable $x$ is impossible to clear. Indeed, there is a non-illuminated path connecting its A-side with its B-side, passing through the B-side of $\ell_1$, the bridge, and the B-side of $\ell_2$. Since $\ell_1$ and $\ell_2$ correspond to incoherent literals, their B-sides are connected to opposite sides of the same variable gadget, by construction.

Summarizing, all the variable gadgets are initially contaminated. In order to clear some of them, a schedule must first select a literal guard from each clause gadget and put it on its A-side. While that position is maintained, there is at least one variable gadget that is impossible to clear. As soon as one literal guard is moved, all the variable gadgets get recontaminated again. It follows that the variable gadgets can never be all clear at the same time, and in particular the polyhedron is unsearchable.

We gave a polynomial time reduction from \TSAT\ to \TSSP\ whose generated numerical coordinates are polynomially bounded in size, hence \TSSP\ is strongly \NP-hard.\end{proof}

\subsubsection*{Optimization problems}
Obviously, the previous theorem implies that the problems of minimizing search time and minimizing \emph{total angular movement} are both \NP-hard to approximate. The first problem stays \NP-hard even when restricted to searchable instances, as proved in \cite{viglietta}. By further inspecting the construction given in that paper, it's clear that the problem does not even have a \PTAS, unless $\mbox{\P} =\mbox{\NP}$. Indeed, satisfiable boolean formulas are transformed into polyhedra searchable in $3$ seconds, while the unsatisfiable ones are transformed into polyhedra that are unsearchable in $3+\varepsilon$ seconds, for a suitable small-enough $\varepsilon > 0$. On the other hand, similar results can be obtained also for the problem of minimizing total angular movement. There are several ways to rearrange the links in the construction employed in Theorem~\ref{hard}, so that the unsatisfiable boolean formulas are mapped into polyhedra that are indeed searchable, but only with very demanding schedules.

\subsection{\PSPACE-hardness of partial searchability}
Here we introduce a slightly generalized problem: suppose the guards have to clear only a given subregion of the polyhedron, while the rest may remain contaminated. In particular, we stipulate that the \emph{target area} that needs to be cleared is expressed as a ball, whose center and radius are given as input along with the polyhedron and the multiset of guards. We call the resulting problem \textsc{3-dimensional Partial Searchlight Scheduling Problem} (\TRSSP).
\begin{definition}[\TRSSP]\emph{\TRSSP}\ is the problem of deciding if the guards of a given instance of \TSSP\ have a schedule that clears a ball with given center and radius.\end{definition}
The terminology defined in Section~\ref{model} for \TSSP\ extends straightforwardly to \TRSSP.

Next we are going to prove that \TRSSP\ is strongly \PSPACE-hard, even restricted to orthogonal polyhedra. To do so, we give a reduction from the \emph{edge-to-edge} problem for \emph{AND/OR constraint graphs} in the \emph{nondeterministic constraint model} described in~\cite{ncl}.

\subsubsection*{Nondeterministic constraint logic machines}
Consider an undirected 3-connected 3-regular planar graph, whose vertices can be of two types: \emph{AND vertices} and \emph{OR vertices}. Of the three edges incident to an AND vertex, one is called its \emph{output edge}, and the other two are its \emph{input edges}. Such a graph is (a special case of) a \emph{nondeterministic constraint logic machine (NCL machine)}. A \emph{legal configuration} of an NCL machine is an orientation (direction) of its edges, such that:
\begin{itemize}
\item for each AND vertex, either its output edge is directed inward, or both its input edges are directed inward;
\item for each OR vertex, at least one of its three incident edges is directed inward.
\end{itemize}
A \emph{legal move} from a legal configuration to another configuration is the reversal of a single edge, in such a way that the above constraints remain satisfied (i.e., such that the resulting configuration is again legal).

Given an NCL machine with two \emph{distinguished edges} $e_a$ and $e_b$, and a \emph{target orientation} for each, we consider the problem of deciding if there are legal configurations $A$ and $B$ such that $e_a$ has its target orientation in $A$, $e_b$ has its target orientation in $B$, and there is a sequence of legal moves from $A$ to $B$. In a sequence of moves, the same edge may be reversed arbitrarily many times. We call this problem \textsc{Edge-to-Edge for Nondeterministic Constraint Logic machines} (\EENCL).

A proof that \EENCL\ is \PSPACE-complete is given in~\cite{ncl}, by a reduction from \textsc{Quantified Boolean Formulas}. Based on that reduction, we may further restrict the set of \EENCL\ instances on which we will be working. Namely, we may assume that $e_a\neq e_b$, and that in no legal configuration both $e_a$ and $e_b$ have their target orientation.

\subsubsection*{Asynchrony}
For our main reduction, it is more convenient to employ an \emph{asynchronous} version of \EENCL. Intuitively, instead of \textquotedblleft instantaneously\textquotedblright\ reversing one edge at a time, we allow any edge to start reversing at any given time, and the \emph{reversal phase} of an edge is not \textquotedblleft atomic\textquotedblright\ and instantaneous, but may take any strictly positive amount of time. It is understood that several edges may be in a reversal phase simultaneously. While an edge is reversing, its orientation is undefined, hence it is not directed toward any vertex. During the whole process, at any time, both the above constraints on AND and OR vertices must be satisfied. We also stipulate that no edge is reversed infinitely many times in a bounded timespan, or else its orientation won't be well-defined in the end. With these extended notions of configuration and move, and with the introduction of \textquotedblleft continuous time\textquotedblright, \EENCL\ is now called \textsc{Edge-to-Edge for Asynchronous Nondeterministic Constraint Logic machines} (\EEANCL).

Despite its asynchrony, such new model of NCL machine has precisely the same power of its traditional synchronous counterpart.
\begin{proposition}\label{asynch}$\EENCL = \EEANCL$.\end{proposition}
\begin{proof}
Obviously $\EENCL \subseteq \EEANCL$, because any sequence of moves in the synchronous model trivially translates into an equivalent sequence for the asynchronous model.

For the opposite inclusion, we show how to \textquotedblleft serialize\textquotedblright\ a legal sequence of moves for an asynchronous NCL machine going from a legal configuration $A$ to configuration $B$ in a bounded timespan, in order to make it suitable for the synchronous model. An asynchronous sequence is represented by a set $S=\{(e_m, s_m, t_m) \mid m\in M\}$, where $M$ is a set of \textquotedblleft edge reversal events\textquotedblright, $e_m$ is an edge with a reversal phase starting at time $s_m$ and terminating at time $t_m > s_m$. For consistency, no two reversal phases of the same edge may overlap.

Because no edge can be reversed infinitely many times, $S$ must be finite. Hence we may assume that $M=\{1, \cdots, n\}$, and that the moves are sorted according to the (weakly increasing) values of $s_m$, i.e., $1\leqslant m < m' \leqslant n \implies s_m \leqslant s_{m'}$. Then we consider the serialized sequence $S'=\{(e_m, m, m) \mid m\in M\}$, and we claim that it is valid for the synchronous model, and that it is equivalent to $S$.

Indeed, each move of $S'$ is instantaneous and atomic, no two edges reverse simultaneously, and every edge is reversed as many times as in $S$, hence the final configuration is again $B$ (provided that the starting configuration is $A$). We still have to show that every move in $S'$ is legal. Let us do the first $m$ edge reversals in $S'$, for some $m\in M$, starting from configuration $A$, and reaching configuration $C$. To prove that $C$ is also legal, consider the configuration $C'$ reached in the asynchronous model at time $s_m$, according to $S$, right when $e_m$ starts its reversal phase (possibly simultaneously with other edges). By construction of $S'$, every edge whose direction is defined in $C'$ (i.e., every edge that is not in a reversal phase) has the same orientation as in $C$. It follows that, for each vertex, its inward edges in $C$ are a superset of its inward edges in $C'$. By assumption on $S$, $C'$ satisfies all the vertex constraints, then so does $C$, a fortiori.
\end{proof}
\begin{corollary}\label{eeancl}\EEANCL\ is \PSPACE-complete.\end{corollary}
\begin{proof}
Recall from~\cite{ncl} that \EENCL\ is \PSPACE-complete, and that $\EENCL = \EEANCL$ by Proposition~\ref{asynch}.
\end{proof}

\subsubsection*{Building blocks}
We realize a given NCL machine in terms of an orthogonal polyhedron consisting of three levels, called \emph{basement}, \emph{floor} and \emph{attic}. The floor contains the actual AND/OR constraint graph, arranged as an orthogonal plane graph, and is completely unsearchable. The attic is reachable from the floor through two \emph{stairs}, and contains the target ball that has to be cleared. The basement level is just a network of \emph{pipes} connecting different parts of the floor to the stairs, whose purpose is to recontaminate the stairs and the attic unless the floor guards actually \textquotedblleft simulate\textquotedblright\ the edges of an NCL machine.

\begin{figure}[ht]
\centering
%\psfrag{a}{A}
\includegraphics[scale=1.25]{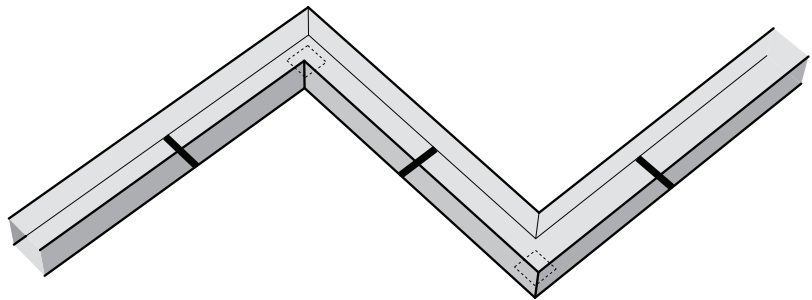}
\caption{An edge, made of three orthogonal subedges.}
\label{figp:bend}
\end{figure}

It is well-known that any 3-regular planar graph can be embedded in the plane as an \emph{orthogonal drawing}. For instance, we can employ the algorithm given in~\cite{planar}, which works with 3-connected 3-regular planar graphs, such as the constraint graphs of our NCL machines. The resulting drawing is orthogonal, in the sense that every edge is a sequence of horizontal and vertical line segments, which will be called \emph{subedges}. For our construction, we turn every subedge into a thin-enough cuboid, then we place a \emph{subedge guard} in the middle of each cuboid, on the bottom face, as depicted in Figure~\ref{figp:bend}. The dotted squares between consecutive subedges are called \emph{trapdoors}, and denote areas that will be attached to pipes and connected to other regions, as described later.

\begin{figure}[ht]
\centering
%\psfrag{a}{A}
\includegraphics[scale=1.25]{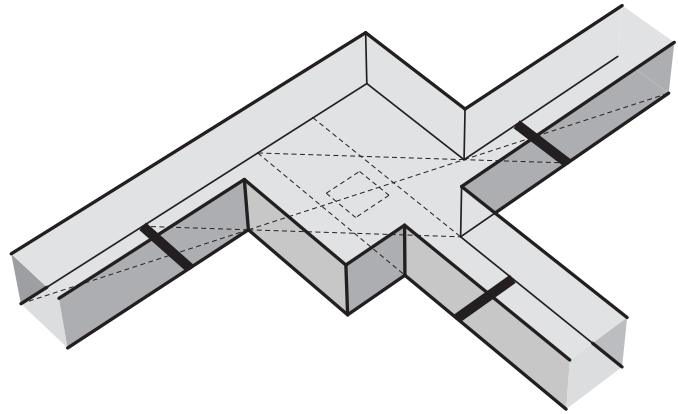}
\caption{An OR vertex.}
\label{figp:or}
\end{figure}

Next we model OR vertices like in Figure~\ref{figp:or}. The three incoming cuboids carrying guards are subedges constructed in the previous paragraph. Again, the dotted square in the middle is a trapdoor that will be attached to pipes. Notice that the trapdoor completely belongs to the visibility region of each of the three guards, as the dotted lines suggest. Moreover, the two subedges coming from opposite directions are displaced, so they do not interfere with each other, in the sense that none of their two guards can see the opposite subedge through the end.

\begin{figure}[ht]
\centering
%\psfrag{a}{A}
\includegraphics[scale=1.25]{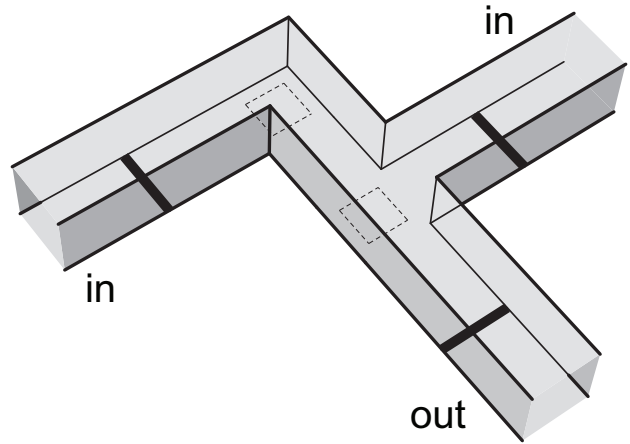}
\caption{An AND vertex.}
\label{figp:and}
\end{figure}

\begin{figure}[ht]
\centering
\subfigure[]{\includegraphics[scale=1.2]{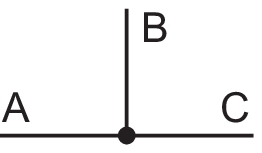}}\qquad \qquad
\subfigure[]{\includegraphics[scale=1.2]{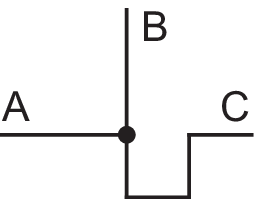}}
\caption{Rearranging the edges of an orthogonal drawing.}
\label{figp:graph}
\end{figure}

We model the AND vertices as shown in Figure~\ref{figp:and}. The output edge is the one whose guard sees both trapdoors, while the guards in the two input edges can see only one trapdoor each. We can always arrange the drawing of our graph by further bending its edges, in such a way that this construction is feasible (i.e., the output edge is located \textquotedblleft between\textquotedblright\ the input edges), as suggested in Figure~\ref{figp:graph}.

Recall that an instance of \EEANCL\ comes with two distinguished edges $e_a$ and $e_b$, each of which is embedded in our construction as a sequence of cuboidal subedges. We select an \emph{internal} subedge $s_a$ of $e_a$ (i.e., not the first, nor the last subedge of $e_a$) and an internal subedge $s_b$ of $e_b$, such that $s_a$ and $s_b$ run in two orthogonal directions. If no such subedges exist in our construction, we can further subdivide $e_a$ and $e_b$ into more small-enough \textquotedblleft redundant\textquotedblright\ subedges, in order to obtain internal ones running in the desired directions. Recall also that both $e_a$ and $e_b$ come in \EEANCL\ with a target orientation, which we want them to reach in order to solve the problem instance. Such target orientation is therefore naturally \textquotedblleft inherited\textquotedblright\ by $s_a$ and $s_b$, too.

\begin{figure}[ht]
\centering
\psfrag{a}{$\ell_a$}
\psfrag{b}{${}_\alpha$}
\includegraphics[scale=1.75]{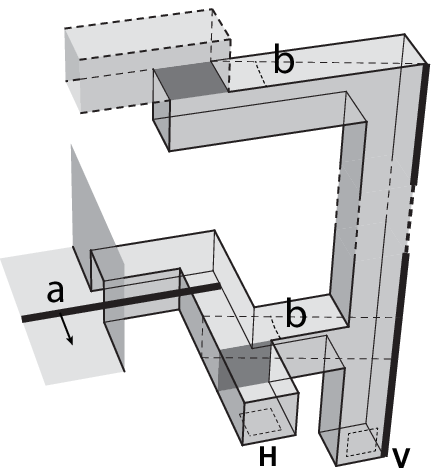}
\caption{A stair.}
\label{figp:stair}
\end{figure}

Let $\ell_a$ and $\ell_b$ be the two subedge guards lying in $s_a$ and $s_b$, respectively. We add two stairs going up to the attic, one for $\ell_a$ and one for $\ell_b$. Figure~\ref{figp:stair} shows a close-up of one end of $\ell_a$, where we have attached a stair (the two \textquotedblleft incomplete\textquotedblright\ rectangles represent faces of $s_a$). Observe that, besides adding the polyhedral model of the stair to $s_a$, we also extend $\ell_a$ to the stair itself. The arrow attached to $\ell_a$ indicates the target orientation of $s_a$, inherited from the instance of \EEANCL. There are two trapdoors, depicted as horizontal and vertical dotted squares. The horizontal trapdoor (marked as H in the picture) lies on the bottom face of an enclosed cuboidal region called \emph{alcove}, whose opening is indicated by a darker vertical square. The alcove completely belongs to $\mathcal V(\ell_a)$, and it can also be capped by the vertical guard showed in the picture, called \emph{stair guard}. While the stair guard caps the alcove, it also covers the vertical trapdoor (marked as V). On the other hand, $\ell_a$ is able to cover the horizontal trapdoor. The top cuboid with dotted edges belongs to the attic, and is not considered part of the stair. The darker horizontal square that separates the attic from the stair is called \emph{attic entrance}.

A similar construction is then repeated for $\ell_b$, and analogous remarks hold.

\begin{figure}[ht]
\centering
%\psfrag{a}{A}
\includegraphics[scale=1.3]{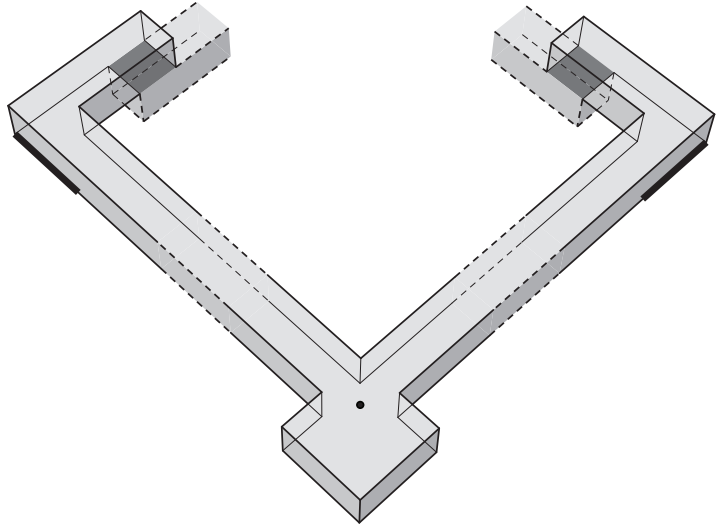}
\caption{The attic.}
\label{figp:attic}
\end{figure}

The whole attic is illustrated in Figure~\ref{figp:attic}, where the darker squares denote the two attic entrances mentioned in the above paragraph, and the underlying cuboids with dotted edges belong to the stairs. A long L-shaped \emph{corridor}, made of two orthogonal \emph{branches}, connects the two entrances, and each entrance can be covered by an \emph{attic guard} located at one end of the corridor. The small dot in the picture, in the middle of the corridor, denotes the target ball that has to be cleared by the guards. There is also a \emph{widening} on one side of the corridor, which is not fully visible to the guards, and hence is a perpetual source of recontamination.

\begin{figure}[ht]
\centering
%\psfrag{a}{A}
\includegraphics[scale=1]{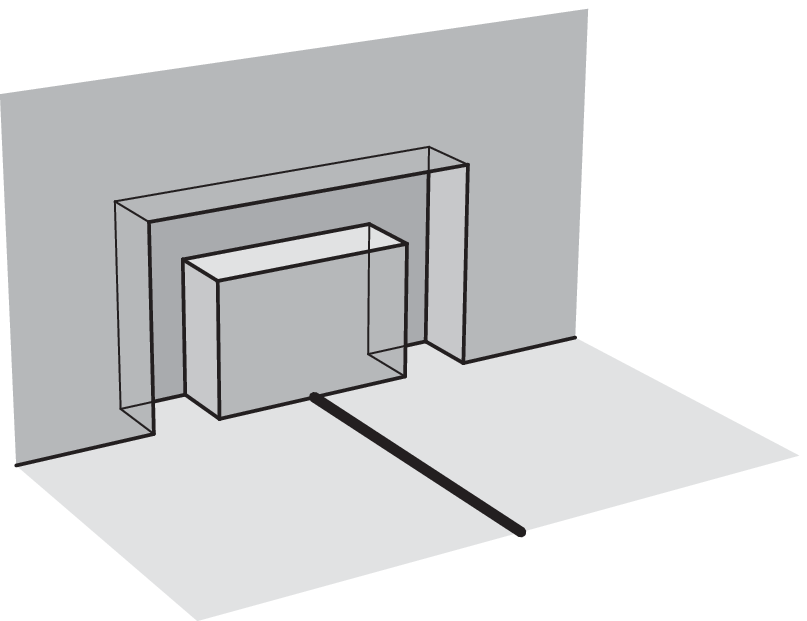}
\caption{A close-up of a subedge guard, with its groove.}
\label{figp:groove}
\end{figure}

To make sure that the floor is really unsearchable, we add a \emph{groove} all around an end of each subedge guard, including $\ell_a$ and $\ell_b$, like in Figure~\ref{figp:groove}. Every groove is always a source of recontamination, because some parts of it can never be seen by any guard, and cannot be isolated from the rest of the floor, either.

\begin{figure}[ht]
\centering
\psfrag{a}{$\ell_1$}
\psfrag{b}{$\ell_2$}
\subfigure[]{\label{figp:pipesa}\includegraphics[scale=0.9]{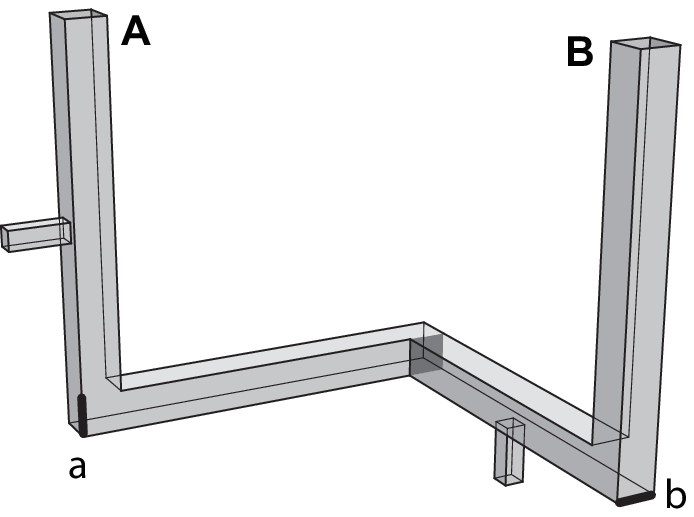}}\qquad \quad
\subfigure[]{\label{figp:pipesb}\includegraphics[scale=0.9]{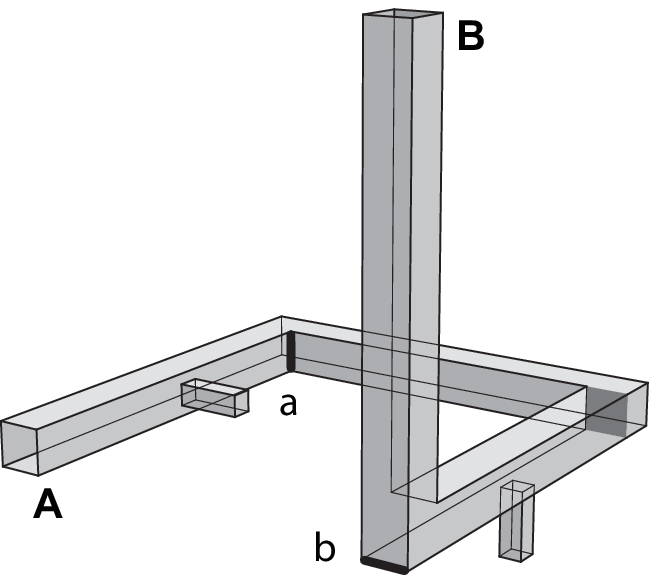}}
\caption{Two pipes. The pipe in \subref{figp:pipesa} connects two horizontal trapdoors, the pipe in \subref{figp:pipesb} connects a vertical and a horizontal trapdoor.}
\label{figp:pipes}
\end{figure}

Finally, at the basement level, we add pipes, i.e., twisted chains of very thin cuboids, to connect pairs of trapdoors. We connect each trapdoor in each stair to every trapdoor in the floor (i.e., the trapdoors in the AND/OR vertices and the trapdoors between subedges). Since there are two different types of trapdoors in the stairs (horizontal and vertical), while the trapdoors in the floor are all horizontal, we need two types of pipes, as Figure~\ref{figp:pipes} suggests. The end of a pipe that is marked as A goes into a stair trapdoor, whereas the end marked as B goes into a floor trapdoor. Observe that a pipe shaped like in Figure~\ref{figp:pipesb} can always connect a vertical stair trapdoor with any floor trapdoor, except when the latter lies exactly \textquotedblleft behind\textquotedblright\ the former, and the lower part of the stair gets in the way. This can be easily prevented, for example by constructing a \textquotedblleft thinner\textquotedblright\ version of the stair itself, such that no floor trapdoor lies completely behind the lower part of the stair (i.e., the part with the vertical trapdoor). In general, all pipes lie below the floor level, and their mutual intersections can be resolved by shrinking them, as already discussed in Subsection~\ref{nphard} for links.

Notice that two very small extra cuboids are attached to each pipe: these are called \emph{pits}. Pits are unsearchable regions, but each pit can indeed be capped by a \emph{pipe guard}. Such guards are also responsible for clearing the rest of the pipe, when both its ends A and B are capped by external searchplanes (belonging to a stair guard and to a subedge guard, respectively). Thus, by \emph{clearing a pipe} we will mean clearing its chain of four bigger cuboids connecting A with B, disregarding the two pits.

\begin{lemma}\label{pipe1}While both its A and B ends are completely illuminated by external guards, a pipe can be cleared.\end{lemma}
\begin{proof}
A similar schedule works for both types of pipe. Referring to Figure~\ref{figp:pipes}, guard $\ell_1$ covers the darker square, allowing $\ell_2$ to clear two of the four cuboids. Then $\ell_2$ keeps capping its pit, while $\ell_1$ sweeps the remaining two cuboids, and caps the other pit.
\end{proof}

On the other hand, an uncapped pipe acts as a \textquotedblleft one-way recontaminator\textquotedblright\ from B to A.

\begin{lemma}\label{pipe2}If neither A nor B is illuminated by external guards, then A is (partly) contaminated.\end{lemma}
\begin{proof}
If $\ell_1$ does not cap its pit, then it contaminates A. If $\ell_1$ caps its pit but $\ell_2$ does not, then the second pit contaminates A. If both guards cap both pits, then B contaminates A.
\end{proof}

\subsubsection*{Reduction}
\begin{theorem}\label{trssp}\TRSSP\ is strongly \PSPACE-hard, even restricted to orthogonal polyhedra.\end{theorem}
\begin{proof}
We give a reduction from \EEANCL\ to \TRSSP, by proving that the target ball in the above construction is clearable if and only if the two distinguished edges $e_a$ and $e_b$ can be oriented in their target directions one after the other, by a legal sequence of asynchronous moves. Observe that our construction is obviously a polyhedron (it is indeed connected) whose vertices' numerical coordinates may be chosen to be polynomially bounded in size, with respect to the size of the constraint graph. Moreover, the orthogonal drawing of the constraint graph can be obtained in linear time, as explained in~\cite{planar}.

\paragraph{Positive instances.}
Suppose that the given instance of \EEANCL\ is solvable. Then there exists a legal sequence $S$ of asynchronous moves that, starting from a configuration $A$ in which $e_a$ is in its target direction, ends in a configuration $B$ in which $e_b$ is in its target direction. By the assumptions we made on constraint graphs, $e_a\neq e_b$ and both $e_a$ and $e_b$ are reversed by $S$.

We start by \textquotedblleft replicating\textquotedblright\ configuration $A$ on the subedge guards in our construction: if an edge has an orientation in $A$, then all its subedges in the drawing inherit its orientation, and all the corresponding subedge guards are oriented accordingly. As a result, every trapdoor in the floor (not the four trapdoors in the stairs) is covered by a guard, since $A$ is a legal configuration. Indeed, the structure of the AND/OR vertices that we built implies that the NCL constraints on a vertex are satisfied if and only if all the trapdoors in its polyhedral model are covered. Also the trapdoors between subedges happen to be covered, because all the guards in a subedge chain are oriented in the same \textquotedblleft direction\textquotedblright. Incidentally, $\ell_a$ covers the horizontal trapdoor in its corresponding stair, as well. Next we cover the vertical trapdoors in both stairs with the stair guards, thus incidentally also capping both alcoves. Finally, we cover both attic entrances with the two attic guards.

In order to clear the target ball in the attic, our schedule proceeds as follows.
\begin{enumerate}
\item Clear the pipes that are attached to the stair trapdoors corresponding to $s_a$. This is feasible by Lemma~\ref{pipe1}, because both ends of such pipes are capped.
\item\label{st2} Replicate all the moves of $S$, with the correct timing. If a move reverses edge $\{u,v\}$ by turning it from vertex $u$ toward vertex $v$, we first reverse the guard corresponding to the subedge incident to $u$, then we reverse all the other subedge guards one by one in order, until we reverse the last guard in the chain, whose subedge is incident to $v$. By doing so, no trapdoor is ever uncovered, hence no pipe is recontaminated. Moreover, both $\ell_a$ and $\ell_b$ reverse during the process, incidentally clearing both alcoves, which are still capped by the stair guards.
\item Clear the pipes attached to the stair trapdoors corresponding to $s_b$, again by Lemma~\ref{pipe1}. Indeed, after Step~\ref{st2}, these pipes are all capped.
\item\label{st3} Clear the regions underlying the two attic entrances, by turning both stair guards by $\alpha$ (refer to Figure~\ref{figp:stair}). Angle $\alpha$ is such that, in the end, both attic entrances are separated from the contaminated floor by the illuminated searchplanes of the stair guards. No contamination is possible through the stair trapdoors either, because all the pipes are still clear (their B ends are still capped).
\item Turn the two attic guards in concert, until they clear the target ball. No recontamination may occur through the attic entrances, whose underlying regions have effectively been cleared in Step~\ref{st3}, whereas the contaminated widening of the corridor is always separated from the portion of corridor that has been swept.
\end{enumerate}

\paragraph{Negative instances.}
Conversely, suppose that no legal sequence of asynchronous moves solves the given instance of \EEANCL, and let us prove that the target ball in the attic is unclearable.

In order to clear the target ball, both attic guards have to turn in concert, away from the attic entrances. Indeed, just one guard is insufficient to clear anything. On the other hand, the attic guards cannot sweep toward the entrances, because of the unavoidable recontaminations from the widening in the corridor.

Therefore, while the attic guards operate, recontamination has to be avoided from the attic entrances. It follows that both stair guards must keep the attic separated from the unsearchable floor. Referring to Figure~\ref{figp:stair}, each stair guard's angle has to be at least $\alpha$.

When in that position, the stair guards cannot cover the vertical trapdoors, nor cap the alcoves. This means that all the pipes attached to a vertical trapdoor and both alcoves have to be simultaneously clear at some time $t$. According to Lemma~\ref{pipe2}, since the stair guards are not capping the A ends of those pipes, then their B ends have to be capped. In other words, all the trapdoors in the floor (not in the stairs) have to be covered. Equivalently, the subedge guards' orientations must correspond to a legal configuration of an asynchronous NCL machine in the given constraint graph. If the orientations of the subedge guards in the chain corresponding to a same edge do not agree with each other, then that edge is considered in a reversal phase.

Recall that, in any legal configuration of the constraint graph, at least one of the two distinguished edges must be oriented opposite to its target direction (it cannot even be in a reversal phase), hence at least one horizontal trapdoor in a stair (say, the stair attached to $s_a$, without loss of generality) must be uncovered at time $t$. However, its alcove and attached pipes must indeed be clear, which means that $\ell_a$ has capped the pipes for some time, allowing them to clear themselves, and then has left the alcove for the last time at $t'$, with $t'<t$. Between time $t'$ and time $t$, the orientation of the subedge guards must always correspond to a legal configuration of the constraint graph, otherwise the alcove just cleared by $\ell_a$ would be recontaminated by its pipes, again by Lemma~\ref{pipe2}.

Observe that, since at time $t'$ guard $\ell_a$ is not oriented strictly opposite to its (inherited) target direction, then $\ell_b$ must be. But, at time $t$, also the alcove corresponding to $s_b$ must be clear, hence there must be some time $t''$, with $t''<t$, when $\ell_b$ last touched that alcove. Once again, between time $t''$ and time $t$, all the subedge guards must be oriented according to some legal configuration of the contraint graph.

Between time $t'$ and time $t''$ (no matter which comes first), the search schedule must simulate an asynchronous legal sequence of moves between a configuration in which $e_a$ has its target orientation (more appropriately, $e_a$ is in a reversal phase), and one in which $e_b$ does. By hypothesis on the given constraint graph, such sequence of moves does not exist: notice in fact that a sequence of moves from $A$ to $B$ is legal if and only if its reverse sequence from $B$ to $A$ is legal.

This concludes the reduction. As a result, because \EEANCL\ is \PSPACE-complete by Corollary~\ref{eeancl}, \TRSSP\ is strongly \PSPACE-hard.
\end{proof}

Observe that the above construction, regarded as an instance if \TSSP, is not viable, as in Definition~\ref{viable}. As a matter of fact, the presence of grooves around some guards, a widening in the attic's corridor and pits in the pipes makes this polyhedron trivially unsearchable. Building a region that is never clear is certainly an effective way to force the recontamination of other critical regions when certain conditions are met. However, if we want our constructions to be viable instances of \TSSP, perhaps in order to prove its \PSPACE-hardness, no such expedient is of any use, and cleverer tools have to be devised.

\section{Concluding remarks}\label{conclusions}
\subsubsection*{Summary}
We modeled the \textsc{3-dimensional Searchlight Scheduling Problem} as a searching problem in polyhedra, where guards are line segments emanating orientable half-planes of light. We showed that exhaustive guards act as the natural counterparts of boundary guards in the 2-dimensional version of the problem, in that the main positive results about \SSP\ still hold for \TSSP\ restricted to instances with just exhaustive guards. By further exploiting the concept of exhaustive guard, we characterized the searchable instances of \TSSP\ with just one guard. Then we proved that $r^2$ guards are sufficient to search any polyhedron with $r>0$ notches, and that just $r$ guards lying over notches are sufficient to search an orthogonal polyhedron, thus partially settling Conjecture~1 in \cite{viglietta}. We also discussed some methods to speed up the search time by placing additional guards. Next we proved that deciding searchability of a polyhedron is strongly \NP-hard, and briefly discussed the hardness of approximation of two related optimization problems. Finally, we showed that deciding if a given target area of an orthogonal polyhedron is searchable is strongly \PSPACE-hard.

\subsubsection*{Further work}
Several problems remain open. Settling Conjecture~1 in \cite{viglietta} would be the main priority. This has been accomplished for orthogonal polyhedra in Theorem~\ref{orth}, but for general polyhedra it turns out to be a surprisingly deep problem. One way to lower the quadratic bound on the number of guards would be to slightly modify the partition used in Theorem~\ref{heur}. Instead of aiming the guards at the angle bisectors of the notches, we could aim them in any direction, cut the polyhedron with the \textquotedblleft extended\textquotedblright\ searchplanes, and still eliminate all the notches. The advantage would be that, by carefully choosing a plane for each notch that minimizes the intersections with other notches, the overall number of intersections could be significantly less than quadratic. Even if it is still quadratic in the worst case, it is likely much lower on average, with respect to any reasonable probability measure over polyhedra.

In the case of orthogonal polyhedra, the upper bound of $r$ guards given by Theorem~\ref{orth} is also not known to be optimal. In fact, we believe that it can still be lowered by a constant factor.

The search time of the schedules given in Theorems~\ref{heur} and \ref{orth} could be dramatically reduced by clearing several regions in parallel.  For instance, in Theorem~\ref{orth} we could turn in concert two guards whose fences do not bound a same cuboid. Generalizing, we could construct a graph $G$ on the set of notches, with an edge for every pair of notches whose fences bound a same cuboid. Then the search time would be proportional to the chromatic number $\chi(G)$, which is likely sublinear, at least on average.

The complexity of \TSSP\ could be considered also for a restricted set of instances, such as 0-genus polyhedra or orthogonal polyhedra. Moreover, the technique used for Theorem~\ref{hard} could perhaps suggest a way to prove the \NP-hardness of \SSP. Adding new elements to \SSP\ in order to increase its expressiveness, while preserving its 2-dimensional nature, could be an intermediate step. For example, we could introduce \emph{curtains}, i.e., line segments that block visibility but don't block movement. Then we could attempt to replicate the construction of Theorem~\ref{hard} for this extension of \SSP.

Similarly, we may try to modify the construction used in Theorem~\ref{trssp} to prove the \PSPACE-hardness of \TSSP\ as well.

\subsection*{Acknowledgements}
The author wishes to thank J.~O'Rourke for his insightful comments.

This work was supported in part by MIUR of Italy under project AlgoDEEP prot.~2008TFBWL4.

%\newpage
%\nocite{*}

\bibliographystyle{plain}
\bibliography{biblio}

\end{document}